\documentclass[reprint,amsmath,amssymb,aps,prl]{revtex4-1}
\usepackage{graphicx,subfigure,bm,amssymb,amsmath,dcolumn,hyperref}
\usepackage{xcolor,multirow}
\usepackage{mathrsfs}
\usepackage[normalem]{ulem}

\begin{document}
	\newcommand{\Tf}{T_{\rm F}}
	\newcommand{\kf}{k_{\rm F}}
	
	\definecolor{blue}{rgb}{0,0,1}
	\definecolor{red}{rgb}{1,0,0}
	\newcommand{\blue}[1]{\textcolor{blue}{#1}}
	\newcommand{\red}[1]{\textcolor{red}{#1}}
	% \title{{\color{red} Phase Diagram of Two-dimensionl Uniform Electron Gas by} Precursory Cooper Flow
 % % : A Physics Phenomenon and Efficient Numeric Protocol
 %        }
        \title{Precursory Cooper Flow in Ultralow-Temperature Superconductors}
	\author{Pengcheng Hou$^{1}$}
	\thanks{These three authors contributed equally to this paper.}
        \author{Xiansheng Cai$^{2}$}
	\thanks{These three authors contributed equally to this paper.}
	\author{Tao Wang$^{2}$}
	\thanks{These three authors contributed equally to this paper.}
	\author{Youjin Deng$^{1,3,4}$}
	\author{Nikolay V. Prokof'ev$^{2}$}
	\author{Boris V. Svistunov$^{2,5}$}
	\author{Kun Chen$^{6}$}
        \email{kunchen@flatironinstitute.org}
	
	\affiliation{$^{1}$ Department of Modern Physics, University of Science and Technology of China, Hefei, Anhui 230026, China}
	\affiliation{$^{2}$ Department of Physics, University of Massachusetts, Amherst, MA 01003, USA}
	\affiliation{$^{3}$ Hefei National Laboratory, University of Science and Technology of China, Hefei 230088, China}
	\affiliation{$^{4}$ MinJiang Collaborative Center for Theoretical Physics, College of Physics and Electronic Information Engineering, Minjiang University, Fuzhou 350108, China}
	\affiliation{$^{5}$ Wilczek Quantum Center, School of Physics and Astronomy, Shanghai Jiao Tong University, Shanghai 200240, China}
	\affiliation{$^{6}$ Center for Computational Quantum Physics, Flatiron Institute, 162 5th Avenue, New York, New York 10010}
	\date{\today}
 
\begin{abstract}

Superconductivity at low temperature---observed in lithium and bismuth, as well as in various low-density superconductors---calls for developing  reliable theoretical and experimental tools for predicting ultralow critical temperatures,  $T_c$, of Cooper instability 
in a system demonstrating nothing but normal Fermi liquid behavior in a broad range of temperatures below the Fermi energy, $\Tf$. 
Equally important
are controlled predictions of stability in a given Cooper channel. We identify such a protocol within the paradigm of precursory Cooper flow---a universal ansatz describing logarithmically slow temperature evolution of the linear response of the normal state to the pair-creating perturbation. Applying this framework to the two-dimensional uniform electron gas, we reveal a series of exotic superconducting states, pushing controlled theoretical predictions of $T_c$ to the unprecedentedly low scale of $10^{-100} \Tf$. 

\end{abstract}
\maketitle

\textit{Introduction.}\textemdash The conceptual elegance of the Kohn-Luttinger theorem establishing that the Fermi liquid state is unstable in high-angular momentum Cooper channels at low temperatures comes 
at the price of lacking accurate predictions as to which channels get unstable first and at what temperatures. Experimentally, each new discovery of the ultralow-temperature superconductor, be it
lithium~\cite{tuoriniemi_superconductivity_2007} or bismuth~\cite{bismuth}, exhibiting superconductivity at about $0.1\,\rm{mK}$, emerges as a surprise.  The potential for observing analogous phenomena in traditionally non-superconducting metals such as gold, copper, or sodium, as well as in low-density superconductors~\cite{uchoa_superconducting_2007,nandkishore_chiral_2012,christos_correlated_2022},
adds to the scientific intrigue, with no {\it a priori} knowledge 
of what the critical temperature $T_c$ one should expect for a given system.

There is, however, a fundamental reason to expect that the desired answers can be controllably extracted---definitely theoretically and,
hopefully, experimentally as well---from the system's properties in the normal Fermi liquid regime at temperatures much lower than the Fermi energy,  $\Tf$, but still many orders of magnitude higher than $T_c$. Indeed, the (ultra-)low value of $T_c$ is due to the emergent Bardeen-Cooper-Schrieffer (BCS) regime of instability, when the strongly correlated at the ultraviolet level system gets renormalized into the Fermi liquid with a weak effective BCS interaction characterized by a dimensionless negative coupling constant $|g| \ll 1$. The result is an exponentially small critical temperature $T_c \sim \Tf e^{1/g}$. The BCS nature of the transitions implies a rather characteristic temperature evolution of the pair susceptibility, $\chi_0(T)$, defined as the linear response to a static pair-creating perturbation. On the approach to the critical point,  $\chi_0(T)$ should behave as \cite{ferrell_fluctuations_1969,scalapino_pair_1970}
\begin{equation}
\chi_0(T)  \propto 1/\ln\left(T/T_c\right) \qquad  (T\to T_c + 0) \, . 
\label{eq:chi0old}
\end{equation}
Experimental studies across a range of superconductors have validated this prediction using superconductor-superconductor tunnel junctions \cite{anderson_experimental_1970,anderson_LT13,bergeal_pairing_2008,she_observing_2011,koren_observation_2016,maier_pairfield_2019}.

\begin{figure}[t]
\centering
\includegraphics[width=0.95\linewidth]{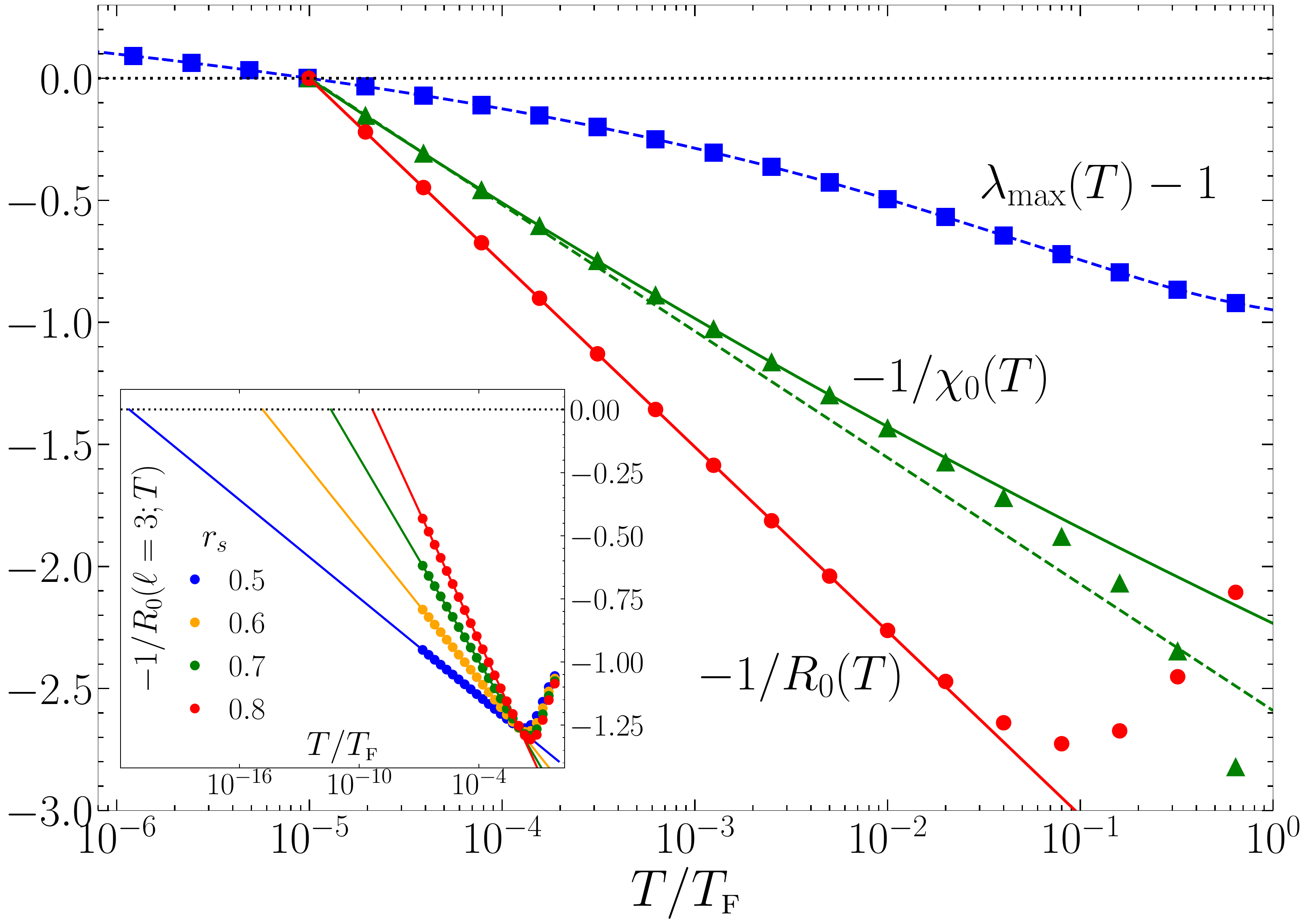}
\caption{Finite-temperature flows of the largest $s$-wave eigenvalue, $\lambda_{\rm max}$, of the standard gap function equation (blue squares connected by line) and linear response functions $\chi_0$ (green triangles) and $R_0$ (red circles) computed for the 2D UEG at $r_s=1.156$. 
The $1/\chi_0$ and $1/R_0$ data are fitted using
Eqs.~\eqref{eq:chi0new}-\eqref{eq:R0}, demonstrating excellent agreement. The
role of the logarithmic numerator in Eq.~\eqref{eq:chi0new} is clearly visible as the difference between the solid green line [fitted with ansatz \eqref{eq:chi0new}] and the dashed green line [fitted with ansatz (\ref{eq:chi0old})].
The inset shows how linear in $\ln T$ scaling of $1/R_0$ is extrapolated to
extract $T_{\rm c}$ in the $f$-channel at several densities. All $1/\chi_0$ and
$1/R_0$ flows undergo qualitative changes near $T/T_{\rm F} \sim 10^{-1}$, where
$T_{\rm F}$ is the Fermi temperature; this sets an energy scale $\Lambda$ below
which the attractive Cooper-channel interaction emerges. 
}
\label{Fig:benchmark} 
\end{figure}

While providing a proof-of-principle result for the idea of extracting $T_c$ from properties 
of the normal state at $T \gg T_c$, relation (\ref{eq:chi0old}) turns out to be rather 
impractical, and sometimes even misleading, when it comes to a controlled quantitative 
analysis of (in)stability in a given pairing channel (for an illustration see the blue curve in Fig.~\ref{Fig:benchmark}).  The reason is that ansatz (\ref{eq:chi0old}) ignores a slowly evolving with temperature logarithmic prefactor (its physical origin is discussed below), making a  na\"ive extrapolation of an apparent linear dependence of $1/\chi_0(T)$ on $\ln T$ from high temperature to the temperature when it is supposed to hit zero very inaccurate, not to mention 
that it may predict finite $T_c$ for a Cooper-stable channel (see a similar discussion in Ref.~\cite{chubukov_implicit_2019}). 

Recently, a numeric method---the so-called implicit renormalization (IR)---allowing one to accurately predict $T_c$ from the field-theoretical properties of the system at $T\gg T_c$  has been proposed in Ref.~\cite{chubukov_implicit_2019} and further developed in Refs.~\cite{cai_superconductivity_2022,wang_origin_2022}, with an application to the model of uniform electron gas. Despite unquestionable success, the IR approach encounters certain technical limitations and lacks a direct connection with what can be measured experimentally. 
%The reformulated gap function (the mathematical object the method deals with) does not have a %direct experimentally observable analog. 
Technical limitations of IR are most pronounced in the vicinity of the ``quantum transition point" (QTP) at which a given channel undergoes a transition from Cooper-stable to Cooper-unstable regime. It is thus crucial to find a complementary to the IR approach and one compatible with experimental protocols.

%%%%%%%%%%%%%%%%%%%%%%%%%%%%%%%%%%%%%%%%%%%%%%%%%%%%%%%%%%%%%%%
%%%%%%%%%%%%%%%%%%%%%%%%%%%%%%%%

In this Letter, we show that the desired solution is provided by nothing but a more accurate than (\ref{eq:chi0new}) but still physically transparent expression for the pair susceptibility.  Specifically, we find that the following (universal to all ultralow-temperature BCS-type superconductors) three-parametric ansatz---the so-called {\it precursory Cooper flow}, or PCF
ansatz---perfectly captures the temperature evolution of $\chi_0$ within a broad temperature range:
\begin{equation}
\chi_0 \, =\,  \frac{c \ln (\Lambda/T)}{1+g \ln (\Lambda/T)} \, +\,  \mathcal O(T) \qquad (T_c < T \ll \Lambda ) \, .
\label{eq:chi0new}
\end{equation}
The non-universal parameters $c$, $g$, and $\Lambda$ are tied to the microscopic properties of the system, with $\Lambda$ being the lowest relevant energy/frequency scale (we set $\hbar=k_{\rm B}=1$), such as $\Tf$, Debye, or plasma frequency. Negative $g$ implies the BCS transition 
at $T_c = \Lambda e^{1/g}$, while $g>0$ implies its absence, with $g=0$  corresponding to QTP.  The logarithmic factor in the numerator---distinguishing (\ref{eq:chi0new}) from (\ref{eq:chi0old})---has the same mathematical origin and, thus, the same expression as the ``Tolmachev's logarithm" in the denominator. However, the two logarithms describe distinctively different physics. The one in the numerator is the pair susceptibility of an {\it ideal} Fermi 
liquid (a system with no coupling in the Cooper channel), while the one in the denominator is responsible for Tolmachev's renormalization of the effective interaction \cite{Tolmachev,simonato_revised_2023}. 
Sharp difference between the stable and unstable regimes develops only 
at $|g| \ln (\Lambda/T) \gg 1$; otherwise, susceptibility increases in both regimes regardless 
of interactions.
In the former case, $\chi_0$ saturates to $c/g$ at temperatures  $T \ll T_* \simeq \Lambda e^{-1/|g|}$. These pair-correlations (diverging as $g\to 0$) play a crucial role in the scenario of strange metal behavior discussed in Ref.~\cite{kapitulnik_colloquium_2019}. %\red{Figure 2(a) demonstrates the temperature-evolution behavior of $\chi_0$ in different regimes.}

For a model with weak momentum-independent interaction, the expression (\ref{eq:chi0new}) is readily obtained by Bethe-Salpeter summation of the Cooper-channel diagrammatic ladder. Far less trivial is our result shedding new light on the previous IR observations \cite{chubukov_implicit_2019,cai_superconductivity_2022,wang_origin_2022} that 
Eq.~(\ref{eq:chi0new}) also works in the case of dynamically screened Coulomb interaction with
complex momentum and frequency dependence of the effective coupling in the Cooper channel.

\begin{figure*}
		\centering
 \includegraphics[width=0.98\linewidth]{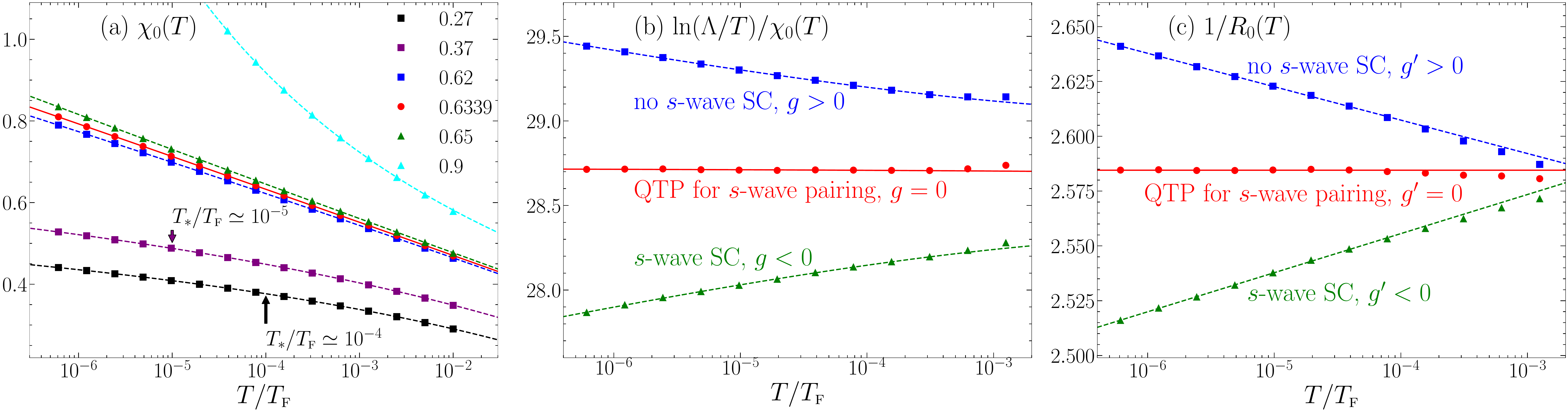}
 \caption{Temperature evolution of the standard pair susceptibility $\chi_0$ and modified pair susceptibility $R_0$ of the 2D uniform electron gas in the $s$-channel for various values of $r_s$. Red circles correspond to QTP $r_s=0.6339$, squares stand for stable regimes ($g>0$), and triangles are used for the unstable regimes ($g<0$). The lines are the fits with the ansatz (\ref{eq:chi0new}) for $\chi_0$ and ansatz (\ref{eq:R0}) for $R_0$. (a) Function $\chi_0 (T)$. For stable regimes, $\chi_0 (T)$ saturates to a constant at $T \ll T_* \simeq \Lambda e^{-1/|g|}$; for unstable regimes, $\chi_0 (T)$ diverges at $T=T_c$; at the QTP, $\chi_0 (T)$ diverges as $T\to 0$. (b) Inverse $\chi_0 (T)$ rescaled with the ideal-gas logarithmic factor. (c) Inverse $R_0$. }
  \label{Fig:lamvsT_rsc} 
	\end{figure*}

In the context of {\it ab initio} calculations employing the PCF methodology, 
we introduce an optimized field-theoretical counterpart of the pair susceptibility,  $R_0$. As opposed to $\chi_0$, the flow of $R_0$ is free of the ``confusing" ideal-Fermi-liquid logarithmic numerator and is characterized by only two parameters: 
\begin{equation}
R_0 (T) = \frac{1}{1 +g' \ln (\Lambda'/T)} + \mathcal O(T) \, .
\label{eq:R0}
\end{equation}
[Consistency with  (\ref{eq:chi0new}) implies $\ln (\Lambda' / \Lambda ) = 1/g' -1/g$.] This yields an exciting opportunity for precise theoretical and numerical determination of $T_c$ and  QTP from normal state calculations using a minimal number of fitting parameters (see Figs.~\ref{Fig:benchmark} and \ref{Fig:lamvsT_rsc}).

\begin{figure}
    \centering
    \includegraphics[width=0.9\linewidth]{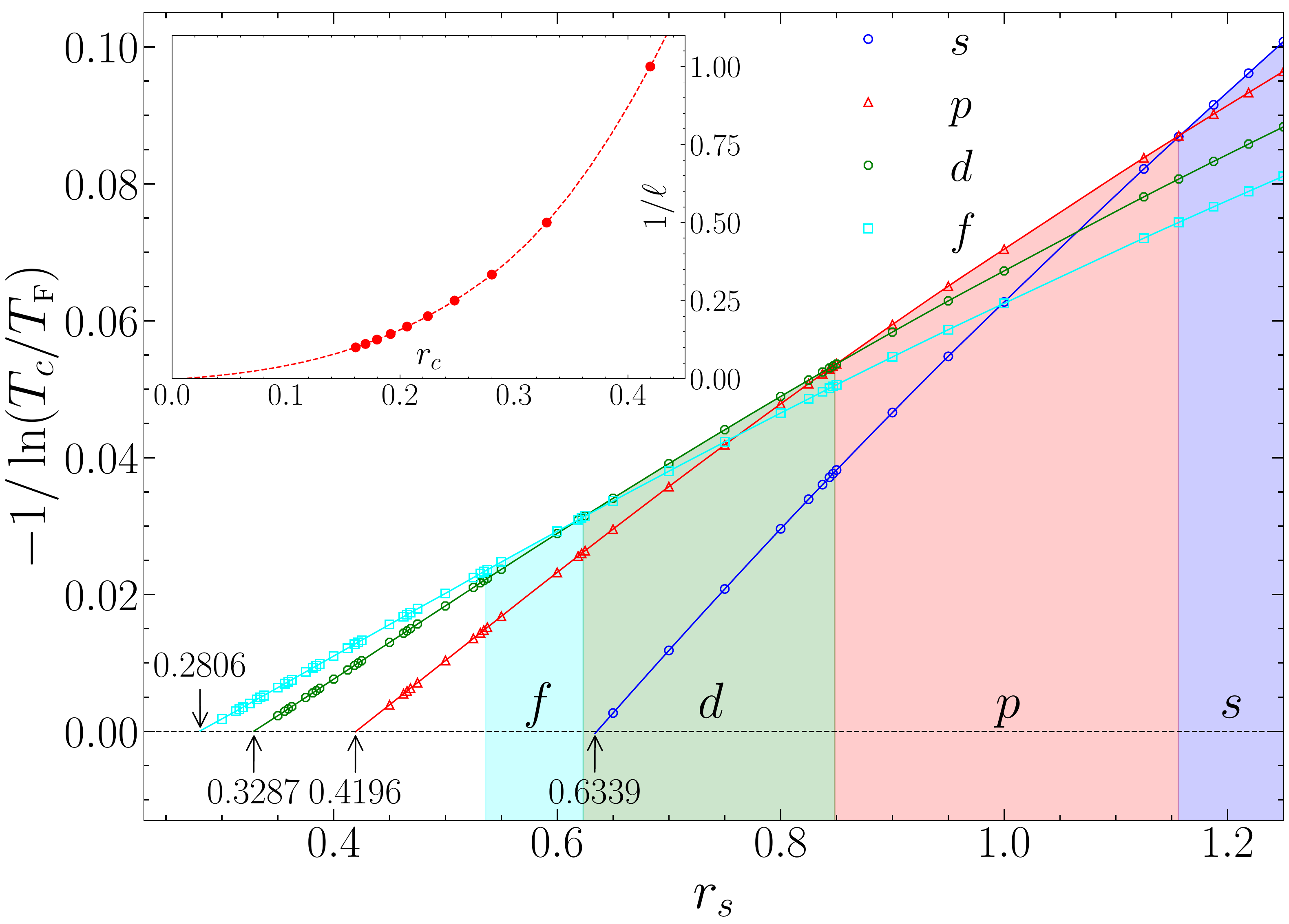}
    \caption{ Superconducting phase diagram of the 2D UEG. 
    For each channel, the line starting at QTP shows the (would-be) critical temperature.
    Critical values of $r_s$ for $\ell$ as large at $10$ are 
    presented in the inset. 
    }
    \label{Fig:phase} 
\end{figure}

With the precise method at hand, we performed a model study of the two-dimensional (2D) uniform electron gas (UEG) in the regime of weak-to-moderate interactions, which is interesting for its intrinsic (no-phonons) superconductivity driven by the dynamically screened Coulomb interaction~\cite{rietschel_role_1983,takada_plasmon_1978,takada_s_1993,galitski_kohn-luttinger_2003,veld2023screening} as opposed to the original Kohn-Luttinger scenario~\cite{kohn_new_1965}. 
Our results (see Fig.~\ref{Fig:phase}) reveal a series of QTPs associated with ultralow-temperature superconducting instabilities that we can resolve down to $10^{-100}\Tf $.  

%%%%%%%%%%%%%%%%%%%%%%%%%%%%%%%%%%%%
%%%%%%%%%%%%%%%%%%%%%%%%%%%%%%%%%%%%%

\textit{Field-theoretical analysis.}\textemdash
Without loss of generality, we study the universal $s$-wave linear response scaling laws in a $d$-dimensional spherically symmetric homogeneous system (other channels and realistic superconductors are discussed in the Supplemental Material~(SM)~\cite{supplement}). The linear response functions \eqref{eq:chi0new} and \eqref{eq:R0} originate from the two-electron Green's function with zero incoming momentum and frequency, $G^{(4)}_{kp}=\Big \langle \mathcal{T} \hat{\psi}_{k\uparrow}^{\dagger} \hat{\psi}_{-k\downarrow}^{\dagger}\hat{\psi}_{-p\downarrow}\hat{\psi}_{p\uparrow} \Big \rangle$, where $\hat{\psi}$/$\hat{\psi}^\dagger$ are the electron annihilation/creation operators. 
We define the shifted momentum-frequency vector as $k = \left(\mathbf{k}-\frac{\mathbf{k}}{|\mathbf{k}|}\kf, \omega_n\right)$ where $\omega_n=(2n+1)\pi T$ is the fermionic Matsubara frequency. Pair susceptibility $\chi_0$
is then the linear response to the static uniform pair-field perturbation
(of unit amplitude), $\chi_0 = \int_k \int_p G^{(4)}_{kp}$, 
where $\int_k \equiv T\sum_n \int \frac{\mathbf{dk}}{(2\pi)^{d}}$. 
The momentum-dependent linear response is defined as 
$R_k = \int_p G^{(4)}_{kp}/(G_kG_{-k})$, where $G_k$ denotes the dressed 
one-electron Green's function.

We start by analyzing the analytic structure of $R_k$ as it follows from the 
self-consistent Bethe-Salpeter equation:
\begin{equation}
R_k= 1 -\int_p \Gamma_{kp} G_p G_{-p}  R_p \,,
%		R_K= \eta(K) -\int dK^\prime \Gamma(K,K^\prime) G^{(2)}(K^\prime) R(K^\prime),
\label{eq:gap}
\end{equation}
where $\Gamma$ is the particle-particle irreducible four-point vertex with zero incoming momentum and frequency. It encodes all effective pairing interactions, such as screened Coulomb potential. 
The second term in the r.h.s. of \eqref{eq:gap} is a sum of ladder diagrams generated by repeated products of $\Gamma$ and $GG$, each carrying its own set of singularities. The finite-temperature cutoff of $GG$ at the Fermi surface is responsible for the logarithmic flow that ultimately leads to BCS instability.  Concurrently, the vertex function $\Gamma$ has singular momentum dependence due to 
incomplete screening of long-range Coulomb interaction at any finite frequency. Possible interplay between the two singularities raises the question of whether the flow
of the pair susceptibility still follows the same law as in the case of short-range interaction.
 
The key observation is that Coulomb singularity does not produce large terms when $\Gamma GG$ is integrated over $p$, and the dominant contribution still comes from the BCS logarithm.
That is, $\int_p \Gamma_{kp}G_pG_{-p}=\tilde{g}_k \ln T + \tilde{f}_k +\mathcal O(T)$, where $\tilde{g}_k$ and $\tilde{f}_k$ are temperature independent and regular in $k$ functions, and the finite-$T$ corrections vanish at least linearly with $T$. Further technical details are provided elsewhere~\cite{longpaper}. This observation
allows one to parameterize $\Gamma GG$ as
\begin{gather}
\label{eq:split}
\Gamma_{kp}G_p G_{-p} \to \left[\tilde{g}_k \ln T + \tilde{f}_k\right] \delta_p + \phi_{kp},
\end{gather}
where $\delta_p = \frac{(2\pi)^{d}\Gamma\left(\frac{d}{2}\right)}{4\pi^{d/2}T}\delta(|\omega_m|-\pi T) \delta(|\mathbf{p}|-\kf)$ (as expected, $\int_p \delta_p =1$), and the regular correction satisfies
$ \int_p \phi_{kp} = \mathcal O(T)$. By incorporating this form into Eq.~\eqref{eq:gap}, we obtain the temperature dependence of the linear response 
(see SM~\cite{supplement}):
\begin{gather}
R_{k} = \frac{1+(f_{k}-f_{0})+(g_{k}-g_{0})\ln T}{1-f_{0} - g_{0}\ln T } + \mathcal{O}(T),
\end{gather}
where $f_k$ and $g_k$ are regular functions representing $\tilde{f}_k$ and $\tilde{g}_k$ renormalized by pair fluctuations. Remarkably, the logarithmic correction in the numerator vanishes in the low-energy limit $k\to 0$, resulting in a simple relation for $R_0 \equiv R_{k\to 0}$ given by Eq.~\eqref{eq:R0} with $g'=g_0$ and $\Lambda' = e^{-f_0/g_0}$. The pair susceptibility, $\chi_0 =\int_k R_k G_k G_{-k}$, on the other hand, involves $R_k$ with finite $k$ and, thus, retains the global logarithmic factor.
%%%%%%%%%%%%%%%%%%%%%%%%%%%%%%%%%%

\textit{Superconductivity in the 2D uniform electron gas.}\textemdash 
In the absence of electron-phonon interaction, superconductivity in this  model is of the emergent BCS type, and the values of $T_{\rm c}$ are supposed to be extremely low. The theory presented above and advanced numerical techniques not only allow us to study many paring channels but also accurately locate QTP when $T_{\rm c}$ goes to zero.

Previous studies \cite{takada_s_1993, cai_superconductivity_2022} revealed the existence of high orbital momentum $\ell$ superconducting states in the 3D UEG in the weak coupling limit when the random phase approximation (RPA) becomes controllably accurate. The pairing comes from dynamic nature screening in Coulomb systems,  as discussed by Rieschel and Sham ~\cite{rietschel_role_1983,buche_superconductivity_1990}, and not from the 
static Kohn-Luttinger mechanism~\cite{kohn_new_1965}. 
In light of the pioneering work by Takada \cite{takada_plasmon_1978}, we expect that the dynamic character of screening will play a crucial role in 2D as well.  

In the weak coupling limit, the RPA vertex function $\Gamma$ has the form
\begin{equation}
	\Gamma_{\mathbf{p}, \omega_{m}}^{\mathbf{k}, \omega_{n}} \approx \frac{V_{\mathbf{p}-\mathbf{k}}}{1+V_{\mathbf{p}-\mathbf{k}} \cdot \Pi_0(\mathbf{p}-\mathbf{k},\omega_m-\omega_n)}\,,
	\end{equation}
 where $V_{\mathbf q}=2\pi e^2/q$ is the bare Coulomb repulsion and 
 $\Pi_0(\mathbf{p}-\mathbf{k},\omega_m-\omega_n)$ is the RPA polarization.
 At this level of accuracy, we consider bare Green's functions $G_{\mathbf k, \omega_n} = 1/(i\omega_n - \mathbf{k}^2/2m +E_{\rm F})$ 
 in Eq.~\eqref{eq:gap}, where $m$ is the electron mass [Fermi temperature can be expressed as $\Tf = 1/(m a_{\rm B}^2 r_s^2)$ where $r_s = \sqrt{2}/(k_{\rm F}a_{\rm B})$ with the Bohr radius $a_{\rm B}$].
    
To efficiently solve Eq.~\eqref{eq:gap} at ultralow temperatures, we employ the discrete Lehmann representation
(DLR)~\cite{kaye_discrete_2022,kaye_libdlr_2022,wang_origin_2022} to 
radically reduce the computational cost of representing linear response and 
Green's functions from ${\mathcal{O}}(1/T\epsilon)$ for uniform grids to $\mathcal{O}[\ln(1/T) \ln(1/\epsilon)]$, where $\epsilon$ is the error tolerance. The codes are available online~\cite{github}.

We performed systematic calculations of $R_0$ for various orbital channels 
$\ell$ and values of $r_s$. For each $\ell$ and $r_s$, we determine the 
transition temperature by extrapolating $1/R_0(T)$ flows to zero 
using the least-squares criterion. Subsequently, for each $\ell$, we extrapolate 
results for $-1/\ln [T_{\rm c}(r_s)/\Tf]$ to zero to reveal critical values of $r_s$.
The phase diagram of competing superconducting states in the 2D UEG 
is shown in Fig.~\ref{Fig:phase}.  

As expected, by increasing density (decreasing $r_s$), one 
suppresses $T_{\rm c}$ in all orbital channels. While all channels are superconducting at $r_s \sim 1$, they successively go through QTPs so that only large $\ell$ channels stay superconducting at small $r_s$. Accordingly, the orbital index of the dominant (highest $T_{\rm c}$) channel also increases with density
(we obtain it from crossing points between the $T_{\rm c}^{(\ell)}(r_s)$ and $T_{\rm c}^{(\ell +1)}(r_s)$ curves). 
Smooth dependence of $ -1/ \ln (T_{\rm c}/\Tf) $ on $r_s$ 
leads to accurate determination of the QTP by linear extrapolation. 
We observe that  $-1/ \ln [T_{\rm c}^{(\ell)}/\Tf] \approx (r_s-r_c^{(\ell)})/c_\ell$, where $c_\ell$ is a dimensionless constant, i.e., in the vicinity of QTP, the transition temperature obeys 
$T_{\rm c}^{(\ell)} = \Tf e^{-1/\lambda_\ell}$~\cite{frank_critical_2007,hainzl_critical_2008},
with $\lambda_\ell = (r_s-r_c^{(\ell)})/c_\ell$.

Data in the inset in Fig.~\ref{Fig:phase} suggest that superconductivity 
in the 2D UEG survives in the high-density limit ($r_s\to 0$) for large enough $\ell$. We emphasize that this outcome cannot be explained by the Kohn-Luttinger mechanism,
because in 2D this mechanism simply does not exist at the RPA level~\cite{baranov_superconductivity_1992}. Similar to the 3D case, 
we are dealing with the consequences of dynamic screening in systems 
with long-range interactions. To explore how interaction range changes 
the picture, we took density corresponding to $r_s=0.8$ (when all channels 
are superconducting) and replaced Coulomb potential with the
Yukawa one. We find that for
screening length $\sim 1/k_{\rm F}$, all channels mentioned in 
Fig.~\ref{Fig:phase} are no longer superconducting. For more details, see SM~\cite{supplement}. 

Finally, we recalculated the phase diagram by accounting for
renormalization of the Green's function within the $G_0W_0$ approximation for the proper self-energy. Qualitative 
(and quantitative at small $r_s$) agreement between phase diagrams presented in the SM~\cite{supplement} and Fig.~\ref{Fig:phase} demonstrates the robustness of our conclusions.
%%%%%%%%%%%%%%%%%%%%%%%%%%%%%%%%%%%%%%%%%%%

\textit{Conclusions and outlook.}\textemdash We have shown that pair susceptibility (linear response to a static spatially uniform pair-creating perturbation) 
of the normal Fermi liquid features universal for all BCS superconductors 
temperature dependence, or ``flow," Eq.~(\ref{eq:chi0new}), irrespective of the emergent pairing mechanism. The ansatz (\ref{eq:chi0new}) applies to both stable and unstable pairing channels. In both cases, the higher-temperature part of the flow is the same, up to small corrections, and  represents singular in the $T\to 0$ limit response of an ideal Fermi liquid. A sharp difference between the stable and unstable cases develops only at exponentially low temperatures: the unstable channel hits finite-temperature singularity at $T_c$ while the stable channel develops 
non-trivial correlations suppressing the zero-temperature singularity. The $T=0$  singularity survives only at the quantum transition points (QTPs) separating the stable and unstable regimes.

Using two-dimensional uniform electron gas as a paradigmatic model of intrinsic superconductivity mediated by dynamic screening of Coulomb interaction, we demonstrated that fitting the normal Fermi liquid data to the ansatz (\ref{eq:chi0new})---and its numeric counterpart (\ref{eq:R0})---allows 
one to accurately predict ultralow critical temperatures of unstable Cooper 
channels and locate QTPs.

We anticipate that our method for controlled quantitative predictions of
ultralow-temperature Cooper instability (or its absence) from finite-temperature 
flows of the linear response to a spatially uniform pair-creating perturbation
can be extended to cases where the perturbation is applied at the 
boundaries of the normal state. This would be the theoretical basis 
for experimental studies of precursory Cooper flows in the normal metallic state 
by using superconductor-normal metal-superconductor tunnel junction setups \cite{anderson_experimental_1970,anderson_LT13,bergeal_pairing_2008,she_observing_2011,koren_observation_2016,maier_pairfield_2019} in the high-temperature 
range $T_c \ll T \ll \Tf$. For metals such as \rm{Cu}, \rm{Au}, and \rm{Na}, where superconductivity at ultra-low temperature remains uncertain, it would be equally informative to observe whether flows (\ref{eq:chi0new}) correspond to 
negative or positive values of $g$ in the $s$-channel.

\textit{Acknowledgements.}\textemdash        
We thank A. Chubukov for valuable discussions. 
P. Hou and Y. Deng were supported by the National Natural Science Foundation of China (under Grant No. 12275263), the Innovation Program for Quantum Science and Technology (under grant no. 2021ZD0301900), and the National Key R\&D Program of China (under Grants No. 2018YFA0306501). N. Prokof'ev, B. Svistunov, and T. Wang were supported by the National Science Foundation under Grant No. DMR-2032077. X. Cai and K. Chen were supported by the Simons Collaboration on the Many Electron Problem. The Flatiron Institute is a division of the Simons Foundation. 

	\bibliographystyle{apsrev}
	\bibliography{references}

\begin{thebibliography}{34}
\expandafter\ifx\csname natexlab\endcsname\relax\def\natexlab#1{#1}\fi
\expandafter\ifx\csname bibnamefont\endcsname\relax
  \def\bibnamefont#1{#1}\fi
\expandafter\ifx\csname bibfnamefont\endcsname\relax
  \def\bibfnamefont#1{#1}\fi
\expandafter\ifx\csname citenamefont\endcsname\relax
  \def\citenamefont#1{#1}\fi
\expandafter\ifx\csname url\endcsname\relax
  \def\url#1{\texttt{#1}}\fi
\expandafter\ifx\csname urlprefix\endcsname\relax\def\urlprefix{URL }\fi
\providecommand{\bibinfo}[2]{#2}
\providecommand{\eprint}[2][]{\url{#2}}

\bibitem[{\citenamefont{Tuoriniemi et~al.}(2007)\citenamefont{Tuoriniemi,
  Juntunen-Nurmilaukas, Uusvuori, Pentti, Salmela, and
  Sebedash}}]{tuoriniemi_superconductivity_2007}
\bibinfo{author}{\bibfnamefont{J.}~\bibnamefont{Tuoriniemi}},
  \bibinfo{author}{\bibfnamefont{K.}~\bibnamefont{Juntunen-Nurmilaukas}},
  \bibinfo{author}{\bibfnamefont{J.}~\bibnamefont{Uusvuori}},
  \bibinfo{author}{\bibfnamefont{E.}~\bibnamefont{Pentti}},
  \bibinfo{author}{\bibfnamefont{A.}~\bibnamefont{Salmela}}, \bibnamefont{and}
  \bibinfo{author}{\bibfnamefont{A.}~\bibnamefont{Sebedash}},
  \bibinfo{journal}{Nature} \textbf{\bibinfo{volume}{447}},
  \bibinfo{pages}{187} (\bibinfo{year}{2007}), ISSN \bibinfo{issn}{0028-0836,
  1476-4687}, \urlprefix\url{https://www.nature.com/articles/nature05820}.

\bibitem[{\citenamefont{Prakash et~al.}(2017)\citenamefont{Prakash, Kumar,
  Thamizhavel, and Ramakrishnan}}]{bismuth}
\bibinfo{author}{\bibfnamefont{O.}~\bibnamefont{Prakash}},
  \bibinfo{author}{\bibfnamefont{A.}~\bibnamefont{Kumar}},
  \bibinfo{author}{\bibfnamefont{A.}~\bibnamefont{Thamizhavel}},
  \bibnamefont{and}
  \bibinfo{author}{\bibfnamefont{S.}~\bibnamefont{Ramakrishnan}},
  \bibinfo{journal}{Science} \textbf{\bibinfo{volume}{355}},
  \bibinfo{pages}{52} (\bibinfo{year}{2017}),
  \urlprefix\url{https://www.science.org/doi/abs/10.1126/science.aaf8227}.

\bibitem[{\citenamefont{Uchoa and
  Castro~Neto}(2007)}]{uchoa_superconducting_2007}
\bibinfo{author}{\bibfnamefont{B.}~\bibnamefont{Uchoa}} \bibnamefont{and}
  \bibinfo{author}{\bibfnamefont{A.~H.} \bibnamefont{Castro~Neto}},
  \bibinfo{journal}{Physical Review Letters} \textbf{\bibinfo{volume}{98}},
  \bibinfo{pages}{146801} (\bibinfo{year}{2007}), ISSN
  \bibinfo{issn}{0031-9007, 1079-7114},
  \urlprefix\url{https://link.aps.org/doi/10.1103/PhysRevLett.98.146801}.

\bibitem[{\citenamefont{Nandkishore et~al.}(2012)\citenamefont{Nandkishore,
  Levitov, and Chubukov}}]{nandkishore_chiral_2012}
\bibinfo{author}{\bibfnamefont{R.}~\bibnamefont{Nandkishore}},
  \bibinfo{author}{\bibfnamefont{L.~S.} \bibnamefont{Levitov}},
  \bibnamefont{and} \bibinfo{author}{\bibfnamefont{A.~V.}
  \bibnamefont{Chubukov}}, \bibinfo{journal}{Nature Physics}
  \textbf{\bibinfo{volume}{8}}, \bibinfo{pages}{158} (\bibinfo{year}{2012}),
  ISSN \bibinfo{issn}{1745-2473, 1745-2481},
  \urlprefix\url{https://www.nature.com/articles/nphys2208}.

\bibitem[{\citenamefont{Christos et~al.}(2022)\citenamefont{Christos, Sachdev,
  and Scheurer}}]{christos_correlated_2022}
\bibinfo{author}{\bibfnamefont{M.}~\bibnamefont{Christos}},
  \bibinfo{author}{\bibfnamefont{S.}~\bibnamefont{Sachdev}}, \bibnamefont{and}
  \bibinfo{author}{\bibfnamefont{M.~S.} \bibnamefont{Scheurer}},
  \bibinfo{journal}{Physical Review X} \textbf{\bibinfo{volume}{12}},
  \bibinfo{pages}{021018} (\bibinfo{year}{2022}), ISSN
  \bibinfo{issn}{2160-3308},
  \urlprefix\url{https://link.aps.org/doi/10.1103/PhysRevX.12.021018}.

\bibitem[{\citenamefont{Ferrell}(1969)}]{ferrell_fluctuations_1969}
\bibinfo{author}{\bibfnamefont{R.~A.} \bibnamefont{Ferrell}},
  \bibinfo{journal}{Journal of Low Temperature Physics}
  \textbf{\bibinfo{volume}{1}}, \bibinfo{pages}{423} (\bibinfo{year}{1969}),
  ISSN \bibinfo{issn}{0022-2291, 1573-7357},
  \urlprefix\url{http://link.springer.com/10.1007/BF00628207}.

\bibitem[{\citenamefont{Scalapino}(1970)}]{scalapino_pair_1970}
\bibinfo{author}{\bibfnamefont{D.~J.} \bibnamefont{Scalapino}},
  \bibinfo{journal}{Phys. Rev. Lett.} \textbf{\bibinfo{volume}{24}},
  \bibinfo{pages}{1052} (\bibinfo{year}{1970}),
  \urlprefix\url{https://link.aps.org/doi/10.1103/PhysRevLett.24.1052}.

\bibitem[{\citenamefont{Anderson and
  Goldman}(1970)}]{anderson_experimental_1970}
\bibinfo{author}{\bibfnamefont{J.~T.} \bibnamefont{Anderson}} \bibnamefont{and}
  \bibinfo{author}{\bibfnamefont{A.~M.} \bibnamefont{Goldman}},
  \bibinfo{journal}{Phys. Rev. Lett.} \textbf{\bibinfo{volume}{25}},
  \bibinfo{pages}{743} (\bibinfo{year}{1970}),
  \urlprefix\url{https://link.aps.org/doi/10.1103/PhysRevLett.25.743}.

\bibitem[{\citenamefont{Anderson et~al.}(1974)\citenamefont{Anderson, Carlson,
  Goldman, and Tan}}]{anderson_LT13}
\bibinfo{author}{\bibfnamefont{J.~T.} \bibnamefont{Anderson}},
  \bibinfo{author}{\bibfnamefont{R.~V.} \bibnamefont{Carlson}},
  \bibinfo{author}{\bibfnamefont{A.~M.} \bibnamefont{Goldman}},
  \bibnamefont{and} \bibinfo{author}{\bibfnamefont{H.-T.} \bibnamefont{Tan}},
  in \emph{\bibinfo{booktitle}{Low {Temperature} {Physics}-{LT} 13: {Volume} 3:
  {Superconductivity}}}, edited by \bibinfo{editor}{\bibfnamefont{K.~D.}
  \bibnamefont{Timmerhaus}}, \bibinfo{editor}{\bibfnamefont{W.~J.}
  \bibnamefont{O’Sullivan}}, \bibnamefont{and}
  \bibinfo{editor}{\bibfnamefont{E.~F.} \bibnamefont{Hammel}}
  (\bibinfo{publisher}{Springer US}, \bibinfo{address}{Boston, MA},
  \bibinfo{year}{1974}), pp. \bibinfo{pages}{709--714}, ISBN
  \bibinfo{isbn}{978-1-4684-2688-5},
  \urlprefix\url{https://doi.org/10.1007/978-1-4684-2688-5_144}.

\bibitem[{\citenamefont{Bergeal et~al.}(2008)\citenamefont{Bergeal, Lesueur,
  Aprili, Faini, Contour, and Leridon}}]{bergeal_pairing_2008}
\bibinfo{author}{\bibfnamefont{N.}~\bibnamefont{Bergeal}},
  \bibinfo{author}{\bibfnamefont{J.}~\bibnamefont{Lesueur}},
  \bibinfo{author}{\bibfnamefont{M.}~\bibnamefont{Aprili}},
  \bibinfo{author}{\bibfnamefont{G.}~\bibnamefont{Faini}},
  \bibinfo{author}{\bibfnamefont{J.~P.} \bibnamefont{Contour}},
  \bibnamefont{and} \bibinfo{author}{\bibfnamefont{B.}~\bibnamefont{Leridon}},
  \bibinfo{journal}{Nature Physics} \textbf{\bibinfo{volume}{4}},
  \bibinfo{pages}{608} (\bibinfo{year}{2008}), ISSN \bibinfo{issn}{1745-2473,
  1745-2481}, \urlprefix\url{http://www.nature.com/articles/nphys1017}.

\bibitem[{\citenamefont{She et~al.}(2011)\citenamefont{She, Overbosch, Sun,
  Liu, Schalm, Mydosh, and Zaanen}}]{she_observing_2011}
\bibinfo{author}{\bibfnamefont{J.-H.} \bibnamefont{She}},
  \bibinfo{author}{\bibfnamefont{B.~J.} \bibnamefont{Overbosch}},
  \bibinfo{author}{\bibfnamefont{Y.-W.} \bibnamefont{Sun}},
  \bibinfo{author}{\bibfnamefont{Y.}~\bibnamefont{Liu}},
  \bibinfo{author}{\bibfnamefont{K.~E.} \bibnamefont{Schalm}},
  \bibinfo{author}{\bibfnamefont{J.~A.} \bibnamefont{Mydosh}},
  \bibnamefont{and} \bibinfo{author}{\bibfnamefont{J.}~\bibnamefont{Zaanen}},
  \bibinfo{journal}{Physical Review B} \textbf{\bibinfo{volume}{84}},
  \bibinfo{pages}{144527} (\bibinfo{year}{2011}), ISSN
  \bibinfo{issn}{1098-0121, 1550-235X},
  \urlprefix\url{https://link.aps.org/doi/10.1103/PhysRevB.84.144527}.

\bibitem[{\citenamefont{Koren and Lee}(2016)}]{koren_observation_2016}
\bibinfo{author}{\bibfnamefont{G.}~\bibnamefont{Koren}} \bibnamefont{and}
  \bibinfo{author}{\bibfnamefont{P.~A.} \bibnamefont{Lee}},
  \bibinfo{journal}{Physical Review B} \textbf{\bibinfo{volume}{94}},
  \bibinfo{pages}{174515} (\bibinfo{year}{2016}), ISSN
  \bibinfo{issn}{2469-9950, 2469-9969},
  \urlprefix\url{https://link.aps.org/doi/10.1103/PhysRevB.94.174515}.

\bibitem[{\citenamefont{Maier and Scalapino}(2019)}]{maier_pairfield_2019}
\bibinfo{author}{\bibfnamefont{T.~A.} \bibnamefont{Maier}} \bibnamefont{and}
  \bibinfo{author}{\bibfnamefont{D.~J.} \bibnamefont{Scalapino}},
  \bibinfo{journal}{npj Quantum Materials} \textbf{\bibinfo{volume}{4}},
  \bibinfo{pages}{30} (\bibinfo{year}{2019}), ISSN \bibinfo{issn}{2397-4648},
  \urlprefix\url{https://www.nature.com/articles/s41535-019-0169-9}.

\bibitem[{\citenamefont{Chubukov et~al.}(2019)\citenamefont{Chubukov,
  Prokof'ev, and Svistunov}}]{chubukov_implicit_2019}
\bibinfo{author}{\bibfnamefont{A.}~\bibnamefont{Chubukov}},
  \bibinfo{author}{\bibfnamefont{N.~V.} \bibnamefont{Prokof'ev}},
  \bibnamefont{and} \bibinfo{author}{\bibfnamefont{B.~V.}
  \bibnamefont{Svistunov}}, \bibinfo{journal}{Physical Review B}
  \textbf{\bibinfo{volume}{100}}, \bibinfo{pages}{064513}
  (\bibinfo{year}{2019}), ISSN \bibinfo{issn}{2469-9950, 2469-9969},
  \urlprefix\url{https://link.aps.org/doi/10.1103/PhysRevB.100.064513}.

\bibitem[{\citenamefont{Cai et~al.}(2022)\citenamefont{Cai, Wang, Prokof'ev,
  Svistunov, and Chen}}]{cai_superconductivity_2022}
\bibinfo{author}{\bibfnamefont{X.}~\bibnamefont{Cai}},
  \bibinfo{author}{\bibfnamefont{T.}~\bibnamefont{Wang}},
  \bibinfo{author}{\bibfnamefont{N.~V.} \bibnamefont{Prokof'ev}},
  \bibinfo{author}{\bibfnamefont{B.~V.} \bibnamefont{Svistunov}},
  \bibnamefont{and} \bibinfo{author}{\bibfnamefont{K.}~\bibnamefont{Chen}},
  \bibinfo{journal}{Physical Review B} \textbf{\bibinfo{volume}{106}},
  \bibinfo{pages}{L220502} (\bibinfo{year}{2022}), ISSN
  \bibinfo{issn}{2469-9950, 2469-9969},
  \urlprefix\url{https://link.aps.org/doi/10.1103/PhysRevB.106.L220502}.

\bibitem[{\citenamefont{Wang et~al.}(2022)\citenamefont{Wang, Cai, Chen,
  Svistunov, and Prokof'ev}}]{wang_origin_2022}
\bibinfo{author}{\bibfnamefont{T.}~\bibnamefont{Wang}},
  \bibinfo{author}{\bibfnamefont{X.}~\bibnamefont{Cai}},
  \bibinfo{author}{\bibfnamefont{K.}~\bibnamefont{Chen}},
  \bibinfo{author}{\bibfnamefont{B.~V.} \bibnamefont{Svistunov}},
  \bibnamefont{and} \bibinfo{author}{\bibfnamefont{N.~V.}
  \bibnamefont{Prokof'ev}}, \emph{\bibinfo{title}{On the {Origin} of {Coulomb}
  {Pseudopotential}: {Two} {Wrongs} {Make} a "{Right}"}}
  (\bibinfo{year}{2022}), \bibinfo{note}{arXiv:2207.05238 [cond-mat]},
  \urlprefix\url{http://arxiv.org/abs/2207.05238}.

\bibitem[{\citenamefont{Bogoliubov et~al.}(1959)\citenamefont{Bogoliubov,
  Tolmachev, and Shirkov}}]{Tolmachev}
\bibinfo{author}{\bibfnamefont{N.}~\bibnamefont{Bogoliubov}},
  \bibinfo{author}{\bibfnamefont{V.}~\bibnamefont{Tolmachev}},
  \bibnamefont{and} \bibinfo{author}{\bibfnamefont{D.}~\bibnamefont{Shirkov}},
  \bibinfo{journal}{New York, Consultants Bureau} p. \bibinfo{pages}{144527}
  (\bibinfo{year}{1959}).

\bibitem[{\citenamefont{Simonato et~al.}(2023)\citenamefont{Simonato,
  Katsnelson, and R\"osner}}]{simonato_revised_2023}
\bibinfo{author}{\bibfnamefont{M.}~\bibnamefont{Simonato}},
  \bibinfo{author}{\bibfnamefont{M.~I.} \bibnamefont{Katsnelson}},
  \bibnamefont{and} \bibinfo{author}{\bibfnamefont{M.}~\bibnamefont{R\"osner}},
  \bibinfo{journal}{Physical Review B} \textbf{\bibinfo{volume}{108}},
  \bibinfo{pages}{064513} (\bibinfo{year}{2023}), ISSN
  \bibinfo{issn}{2469-9950, 2469-9969},
  \urlprefix\url{https://link.aps.org/doi/10.1103/PhysRevB.108.064513}.

\bibitem[{\citenamefont{Kapitulnik et~al.}(2019)\citenamefont{Kapitulnik,
  Kivelson, and Spivak}}]{kapitulnik_colloquium_2019}
\bibinfo{author}{\bibfnamefont{A.}~\bibnamefont{Kapitulnik}},
  \bibinfo{author}{\bibfnamefont{S.~A.} \bibnamefont{Kivelson}},
  \bibnamefont{and} \bibinfo{author}{\bibfnamefont{B.}~\bibnamefont{Spivak}},
  \bibinfo{journal}{Reviews of Modern Physics} \textbf{\bibinfo{volume}{91}},
  \bibinfo{pages}{011002} (\bibinfo{year}{2019}), ISSN
  \bibinfo{issn}{0034-6861, 1539-0756},
  \urlprefix\url{https://link.aps.org/doi/10.1103/RevModPhys.91.011002}.

\bibitem[{\citenamefont{Rietschel and Sham}(1983)}]{rietschel_role_1983}
\bibinfo{author}{\bibfnamefont{H.}~\bibnamefont{Rietschel}} \bibnamefont{and}
  \bibinfo{author}{\bibfnamefont{L.~J.} \bibnamefont{Sham}},
  \bibinfo{journal}{Physical Review B} \textbf{\bibinfo{volume}{28}},
  \bibinfo{pages}{5100} (\bibinfo{year}{1983}), ISSN \bibinfo{issn}{0163-1829},
  \urlprefix\url{https://link.aps.org/doi/10.1103/PhysRevB.28.5100}.

\bibitem[{\citenamefont{Takada}(1978)}]{takada_plasmon_1978}
\bibinfo{author}{\bibfnamefont{Y.}~\bibnamefont{Takada}},
  \bibinfo{journal}{Journal of the Physical Society of Japan}
  \textbf{\bibinfo{volume}{45}}, \bibinfo{pages}{786} (\bibinfo{year}{1978}),
  ISSN \bibinfo{issn}{0031-9015, 1347-4073},
  \urlprefix\url{https://journals.jps.jp/doi/10.1143/JPSJ.45.786}.

\bibitem[{\citenamefont{Takada}(1993)}]{takada_s_1993}
\bibinfo{author}{\bibfnamefont{Y.}~\bibnamefont{Takada}},
  \bibinfo{journal}{Physical Review B} \textbf{\bibinfo{volume}{47}},
  \bibinfo{pages}{5202} (\bibinfo{year}{1993}), ISSN \bibinfo{issn}{0163-1829,
  1095-3795},
  \urlprefix\url{https://link.aps.org/doi/10.1103/PhysRevB.47.5202}.

\bibitem[{\citenamefont{Galitski and
  Das~Sarma}(2003)}]{galitski_kohn-luttinger_2003}
\bibinfo{author}{\bibfnamefont{V.~M.} \bibnamefont{Galitski}} \bibnamefont{and}
  \bibinfo{author}{\bibfnamefont{S.}~\bibnamefont{Das~Sarma}},
  \bibinfo{journal}{Physical Review B} \textbf{\bibinfo{volume}{67}},
  \bibinfo{pages}{144520} (\bibinfo{year}{2003}), ISSN
  \bibinfo{issn}{0163-1829, 1095-3795},
  \urlprefix\url{https://link.aps.org/doi/10.1103/PhysRevB.67.144520}.

\bibitem[{\citenamefont{in~'t Veld et~al.}(2023)\citenamefont{in~'t Veld,
  Katsnelson, Millis, and R\"osner}}]{veld2023screening}
\bibinfo{author}{\bibfnamefont{Y.}~\bibnamefont{in~'t Veld}},
  \bibinfo{author}{\bibfnamefont{M.~I.} \bibnamefont{Katsnelson}},
  \bibinfo{author}{\bibfnamefont{A.~J.} \bibnamefont{Millis}},
  \bibnamefont{and} \bibinfo{author}{\bibfnamefont{M.}~\bibnamefont{R\"osner}},
  \emph{\bibinfo{title}{Screening induced crossover between phonon- and
  plasmon-mediated pairing in layered superconductors}} (\bibinfo{year}{2023}),
  \eprint{arXiv:2303.06220}.

\bibitem[{\citenamefont{Kohn and Luttinger}(1965)}]{kohn_new_1965}
\bibinfo{author}{\bibfnamefont{W.}~\bibnamefont{Kohn}} \bibnamefont{and}
  \bibinfo{author}{\bibfnamefont{J.~M.} \bibnamefont{Luttinger}},
  \bibinfo{journal}{Physical Review Letters} \textbf{\bibinfo{volume}{15}},
  \bibinfo{pages}{524} (\bibinfo{year}{1965}), ISSN \bibinfo{issn}{0031-9007},
  \urlprefix\url{https://link.aps.org/doi/10.1103/PhysRevLett.15.524}.

\bibitem[{sup()}]{supplement}
\bibinfo{note}{See Supplemental Material for additional information}.

\bibitem[{lon()}]{longpaper}
\bibinfo{note}{Paper in preparation by the authors}.

\bibitem[{\citenamefont{B\"uche and
  Rietschel}(1990)}]{buche_superconductivity_1990}
\bibinfo{author}{\bibfnamefont{T.}~\bibnamefont{B\"uche}} \bibnamefont{and}
  \bibinfo{author}{\bibfnamefont{H.}~\bibnamefont{Rietschel}},
  \bibinfo{journal}{Physical Review B} \textbf{\bibinfo{volume}{41}},
  \bibinfo{pages}{8691} (\bibinfo{year}{1990}), ISSN \bibinfo{issn}{0163-1829,
  1095-3795},
  \urlprefix\url{https://link.aps.org/doi/10.1103/PhysRevB.41.8691}.

\bibitem[{\citenamefont{Kaye et~al.}(2022{\natexlab{a}})\citenamefont{Kaye,
  Chen, and Parcollet}}]{kaye_discrete_2022}
\bibinfo{author}{\bibfnamefont{J.}~\bibnamefont{Kaye}},
  \bibinfo{author}{\bibfnamefont{K.}~\bibnamefont{Chen}}, \bibnamefont{and}
  \bibinfo{author}{\bibfnamefont{O.}~\bibnamefont{Parcollet}},
  \bibinfo{journal}{Physical Review B} \textbf{\bibinfo{volume}{105}},
  \bibinfo{pages}{235115} (\bibinfo{year}{2022}{\natexlab{a}}), ISSN
  \bibinfo{issn}{2469-9950, 2469-9969},
  \urlprefix\url{https://link.aps.org/doi/10.1103/PhysRevB.105.235115}.

\bibitem[{\citenamefont{Kaye et~al.}(2022{\natexlab{b}})\citenamefont{Kaye,
  Chen, and Strand}}]{kaye_libdlr_2022}
\bibinfo{author}{\bibfnamefont{J.}~\bibnamefont{Kaye}},
  \bibinfo{author}{\bibfnamefont{K.}~\bibnamefont{Chen}}, \bibnamefont{and}
  \bibinfo{author}{\bibfnamefont{H.~U.} \bibnamefont{Strand}},
  \bibinfo{journal}{Computer Physics Communications}
  \textbf{\bibinfo{volume}{280}}, \bibinfo{pages}{108458}
  (\bibinfo{year}{2022}{\natexlab{b}}), ISSN \bibinfo{issn}{00104655},
  \urlprefix\url{https://linkinghub.elsevier.com/retrieve/pii/S0010465522001771}.

\bibitem[{git()}]{github}
\bibinfo{note}{This is a link to the code of the linear response approach in
  this paper.}, \urlprefix\url{https://github.com/numericalEFT/ElectronGas.jl}.

\bibitem[{\citenamefont{Frank et~al.}(2007)\citenamefont{Frank, Hainzl, Naboko,
  and Seiringer}}]{frank_critical_2007}
\bibinfo{author}{\bibfnamefont{R.~L.} \bibnamefont{Frank}},
  \bibinfo{author}{\bibfnamefont{C.}~\bibnamefont{Hainzl}},
  \bibinfo{author}{\bibfnamefont{S.}~\bibnamefont{Naboko}}, \bibnamefont{and}
  \bibinfo{author}{\bibfnamefont{R.}~\bibnamefont{Seiringer}},
  \bibinfo{journal}{Journal of Geometric Analysis}
  \textbf{\bibinfo{volume}{17}}, \bibinfo{pages}{559} (\bibinfo{year}{2007}),
  ISSN \bibinfo{issn}{1050-6926, 1559-002X},
  \urlprefix\url{http://link.springer.com/10.1007/BF02937429}.

\bibitem[{\citenamefont{Hainzl and Seiringer}(2008)}]{hainzl_critical_2008}
\bibinfo{author}{\bibfnamefont{C.}~\bibnamefont{Hainzl}} \bibnamefont{and}
  \bibinfo{author}{\bibfnamefont{R.}~\bibnamefont{Seiringer}},
  \bibinfo{journal}{Physical Review B} \textbf{\bibinfo{volume}{77}},
  \bibinfo{pages}{184517} (\bibinfo{year}{2008}), ISSN
  \bibinfo{issn}{1098-0121, 1550-235X},
  \urlprefix\url{https://link.aps.org/doi/10.1103/PhysRevB.77.184517}.

\bibitem[{\citenamefont{Baranov et~al.}(1992)\citenamefont{Baranov, Chubukov,
  and Yu.~Kagan}}]{baranov_superconductivity_1992}
\bibinfo{author}{\bibfnamefont{M.~A.} \bibnamefont{Baranov}},
  \bibinfo{author}{\bibfnamefont{A.~V.} \bibnamefont{Chubukov}},
  \bibnamefont{and}
  \bibinfo{author}{\bibfnamefont{M.}~\bibnamefont{Yu.~Kagan}},
  \bibinfo{journal}{International Journal of Modern Physics B}
  \textbf{\bibinfo{volume}{06}}, \bibinfo{pages}{2471} (\bibinfo{year}{1992}),
  ISSN \bibinfo{issn}{0217-9792, 1793-6578},
  \urlprefix\url{https://www.worldscientific.com/doi/abs/10.1142/S0217979292001249}.

\end{thebibliography}


\begin{thebibliography}{1}
\expandafter\ifx\csname natexlab\endcsname\relax\def\natexlab#1{#1}\fi
\expandafter\ifx\csname bibnamefont\endcsname\relax
  \def\bibnamefont#1{#1}\fi
\expandafter\ifx\csname bibfnamefont\endcsname\relax
  \def\bibfnamefont#1{#1}\fi
\expandafter\ifx\csname citenamefont\endcsname\relax
  \def\citenamefont#1{#1}\fi
\expandafter\ifx\csname url\endcsname\relax
  \def\url#1{\texttt{#1}}\fi
\expandafter\ifx\csname urlprefix\endcsname\relax\def\urlprefix{URL }\fi
\providecommand{\bibinfo}[2]{#2}
\providecommand{\eprint}[2][]{\url{#2}}

\bibitem[{\citenamefont{Frye and Efthimiou}(2012)}]{frye_spherical_2012}
\bibinfo{author}{\bibfnamefont{C.}~\bibnamefont{Frye}} \bibnamefont{and}
  \bibinfo{author}{\bibfnamefont{C.~J.} \bibnamefont{Efthimiou}},
  \bibinfo{journal}{arXiv:1205.3548 [hep-th, physics:math-ph]}
  (\bibinfo{year}{2012}), \urlprefix\url{http://arxiv.org/abs/1205.3548}.

\end{thebibliography}
\end{document}

\newcommand{\Tf}{T_{\textsc f}}
\newcommand{\kf}{k_{\textsc f}}
\definecolor{blue}{rgb}{0,0,1}
\definecolor{red}{rgb}{1,0,0}
\newcommand{\blue}[1]{\textcolor{blue}{#1}}
\newcommand{\red}[1]{\textcolor{red}{#1}}

\title{\underline{Supplemental Material:} Precursory Cooper Flow in Ultralow-Temperature Superconductors}
	\author{Pengcheng Hou$^{1}$}
	\thanks{These three authors contributed equally to this paper.}
	\author{Xiansheng Cai$^{2}$}
	\thanks{These three authors contributed equally to this paper.}
	\author{Tao Wang$^{2}$}
	\thanks{These three authors contributed equally to this paper.}
	\author{Youjin Deng$^{1,3,4}$}
	\author{Nikolay V. Prokof’ev$^{2}$}
	\author{Boris V. Svistunov$^{2,5}$}
	\author{Kun Chen$^{6}$}	\email{kunchen@flatironinstitute.org}
 
	\affiliation{$^{1}$ Department of Modern Physics, University of Science and Technology of China, Hefei, Anhui 230026, China}
	\affiliation{$^{2}$ Department of Physics, University of Massachusetts, Amherst, MA 01003, USA}
	\affiliation{$^{3}$ Hefei National Laboratory, University of Science and Technology of China, Hefei 230088, China}
	\affiliation{$^{4}$ MinJiang Collaborative Center for Theoretical Physics, College of Physics and Electronic Information Engineering, Minjiang University, Fuzhou 350108, China}
	\affiliation{$^{5}$ Wilczek Quantum Center, School of Physics and Astronomy, Shanghai Jiao Tong University, Shanghai 200240, China}
	\affiliation{$^{6}$ Center for Computational Quantum Physics, Flatiron Institute, 162 5th Avenue, New York, New York 10010}

\maketitle

\section{Angular momentum decomposition in two and higher dimensions}
In an isotropic system, the particle-particle four-point vertex with zero incoming momentum and frequency $\Gamma_{\mathbf p,\omega_m}^{\mathbf k,\omega_n}$ only depends on the angle between $\mathbf{k}$ and $\mathbf{p}$ ($\theta_{\hat{kp}} \equiv \theta_{\hat k} - \theta_{\hat p}$), their moduli, and the Matsubara-frequencies difference $\omega_n-\omega_m$. This allows us to simplify the analysis by projecting $\Gamma$ onto different orbital channels, $\ell$. In this section, the frequency variables and Mataubara summation are omitted for brevity as they do not affect the decomposition.

Considering the Bethe-Salpeter equation for the linear response function $R$,
\begin{equation}
     R_{\mathbf{k}}= \eta_{\mathbf{k}} -\int  \frac{d\mathbf{p}}{(2\pi)^d} \Gamma_{\mathbf{p}}^{\mathbf{k}} F_{\mathbf{p}} \,,
% R_K = \eta_K - \int dP \Gamma_{K,P} F_P \,,
\label{eq:BS}
\end{equation}
where $\eta_{\mathbf k}$ is the sourced term and $F_{\mathbf{p}}=G_{\mathbf{p}} G_{-\mathbf{p}}  R_{\mathbf{p}}$, we can project it onto decoupled orbital channels $\ell$ via an expansion. After the projection to the angular momentum sector, $R$ and $\Gamma$ are solely dependent on the momentum amplitudes, allowing a unified treatment of generic dimension. 

In two dimensions, the vertex function can be expressed as a Fourier expansion
\begin{equation}
\begin{aligned}
	\Gamma_{|\mathbf{k}-\mathbf{p}|}&= \frac{\Gamma^{(0)}}{2\pi} + \sum_{\ell=1}^{\infty} \Gamma^{(\ell)}_{k, p} \frac{\cos\ell \theta_{\hat{kp}}}{\pi}   \,, \\
	\Gamma^{(\ell)}_{k, p}&=\int_{0}^{2\pi} d\theta_{\hat{kp}}  \Gamma\left(\sqrt{k^2+p^2-2kp\cos \theta_{\hat{kp}}}\right) \cos \ell \theta_{\hat{kp}} \,.
\end{aligned}
  \label{eq:orbital}
\end{equation}
Apart from the $\Gamma$ expansion, Eq.~\eqref{eq:orbital}, 
we also need to expand functions that depend only on one momentum variable:
\begin{equation}
\begin{aligned}
O_{\mathbf{k}} &= \frac{O^{(0)}_k}{2\pi}+ \sum_{\ell=1}^{\infty} O^{(\ell)}_k \frac{\cos\ell \theta_{\hat k}}{\pi} +\overline{O}^{(\ell)}_k \frac{\sin\ell \theta_{\hat k}}{\pi} \,, 
% \eta_K &= \frac{\eta^{(0)}_K}{2\pi}+ \sum_{\ell=1} \eta^{(\ell)}_K \frac{\cos\ell \theta_{\hat k}}{\pi} +\overline{\eta}^{(\ell)}_K \frac{\sin\ell \theta_{\hat k}}{\pi}\,,
\end{aligned}
\end{equation}
where $O$ represents $R$, $F$, or $\eta$, $O^{(\ell)}_k =\int_{0}^{2\pi} d\theta_{\hat k} \cos\ell \theta_{\hat k} O_{\mathbf{k}}$, and $\overline{O}^{(\ell)}_k =\int_{0}^{2\pi} d\theta_{\hat k} \sin\ell \theta_{\hat k} O_{\mathbf{k}}$. 
%The product of two Green's function is isotropic.
%{\color{red} You did not introduce expansion of F and it is not clear why you are not using $R$ everywhere. I would change this}
The convolution term in Eq.\eqref{eq:BS} can also be expanded in Fourier series:
\begin{widetext}
\begin{equation}
\begin{aligned}
&\quad\; \int \frac{d\mathbf p}{(2\pi)^2} \Gamma_{|\mathbf{k}-\mathbf{p}|} F_{\mathbf p} \\
&= \int \frac{d \mathbf p}{(2\pi)^2} 
\left\{
\left( \frac{\Gamma^{(0)}_{k,p}}{2\pi} + 
\sum_{\ell=1}^{\infty} \Gamma^{(\ell)}_{k,p} \frac{\cos \ell \theta_{\hat{kp}}}{\pi}   \right)
\left( \frac{F^{(0)}_p}{2\pi}+\sum_{\ell^\prime=1}^{\infty} F^{(\ell^\prime)}_p \frac{\cos\ell^\prime \theta_{\hat p}}{\pi}  + \overline{F}^{(\ell^\prime)}_p \frac{\sin\ell^\prime \theta_{\hat p}}{\pi}  \right)
\right\}\\
&= \int \frac{pdp}{(2\pi)^2} \left\{ \Gamma^{(0)}_{k,p} \frac{F^{(0)}_p}{2\pi} + \sum_{\ell=1}^{\infty} \Gamma^{(\ell)}_{k,p} \int_0^{2\pi} \frac{d\theta_{\hat p}}{\pi}
\left( \cos\ell \theta_{\hat k}\cos\ell \theta_{\hat p} + \sin\ell \theta_{\hat k}\sin\ell \theta_{\hat p} \right) \sum_{\ell^\prime=1}^{\infty} \left[ F^{(\ell^\prime)}_p \frac{\cos\ell^\prime \theta_{\hat p}}{\pi}  + \overline{F}^{(\ell^\prime)}_p \frac{\sin\ell^\prime \theta_{\hat p}}{\pi} \right] \right\} \\
&= \int \frac{pdp}{(2\pi)^2} \left\{ \Gamma^{(0)}_{k,p} \frac{F^{(0)}_p}{2\pi} + \sum_{\ell=1}^{\infty} \Gamma^{(\ell)}_{k,p} \left[F^{(\ell)}_p \frac{\cos\ell \theta_{\hat k}}{\pi} +\overline{F}^{(\ell)}_p \frac{\sin\ell \theta_{\hat k}}{\pi} \right]\right\} \,,
\end{aligned}
\end{equation}
\end{widetext}
where $F^{(\ell)}_p = G_p G_{-p} R^{(\ell)}_p$ and $R^{(\ell)}_p= \int_{0}^{2\pi} d\theta_{\hat p} \cos\ell \theta_{\hat p} R_{\mathbf p}$ (the product of two Green's functions is isotropic).
Therefore, the linear response function for each channel is given by 
\begin{equation}
    R^{(\ell)}_k %= 1 - \int \frac{p {\rm d}p}{(2\pi)^2} \Gamma^{(\ell)}_{k,p} F^{(\ell)}_p 
    = 1 - \int \frac{p {\rm d}p}{(2\pi)^2} \Gamma^{(\ell)}_{k,p} G_p G_{-p} R^{(\ell)}_p \,.
\label{eq:2DR}
\end{equation}
 Here by setting $\eta^{(\ell)} \equiv 1$ and projecting into channel $\ell$, we break the $U(1)$ symmetry and the rotation symmetry so that the result for different channels can be obtained separately. 

In three or higher dimensions, the angular momentum decomposition can be achieved by a similar approach using expansion in the spherical harmonics. The vertex function can be expressed as an expansion in Legendre polynomials $P_{\ell}(x)$
\begin{equation}  
\begin{aligned}
\Gamma_{|\mathbf k-\mathbf p|} &=\sum_{\ell=0}^{\infty}\frac{N(d,\ell)}{2} \Gamma^{(\ell)}_{k,p} P_{\ell}(x) \,,\\
\Gamma^{(\ell)}_{k,p}&=\int_{-1}^{1} dx  \Gamma\left(\sqrt{k^2+p^2-2kpx} \right) P_{\ell}(x) \,,
\end{aligned}
\end{equation}
where $x=\cos \theta_{\hat{kp}}$ and $N(d,\ell)= \frac{2\ell+d-2}{\ell}  \binom{\ell+d-3}{\ell-1}$ denotes the number of linearly independent Legendre polynomials of degree $\ell$ in $d$ dimensions. Using the addition theorem for spherical harmonics~\cite{frye_spherical_2012}  
\begin{equation}
P_{\ell}(x)=\frac{\Omega_{d}}{N(d,\ell)} \sum_{m=1}^{N(d,\ell)} Y_{\ell m}(\theta_{\hat k}) Y_{\ell m}^{*}(\theta_{\hat p})\,,
\end{equation}
where $\Omega_{d}=(2\pi)^{\frac d 2}/\Gamma(d/2)$ is the solid angle in $d$ dimensions and $Y_{\ell m}(\theta)$ is the spherical harmonic function, the vertex function can be expressed further on as 
\begin{equation}
  \Gamma_{|\mathbf k-\mathbf p|}=\frac{\Omega_d}{2} \sum_{\ell=0}^{\infty} \Gamma^{(\ell)}_{k,p} Y_{\ell 0}(\theta_{\hat k})Y_{\ell 0}^{*}(\theta_{\hat p}) \,.
\end{equation}
Here we set $m=0$ because the decomposed equation is independent of $m$ for a system with rotation symmetry. 
The linear response function can also be decomposed as
\begin{equation}
  R_{\mathbf{k}} = \sum_{\ell=0}^{\infty} R^{(\ell)}_k Y_{\ell 0}(\theta_{\hat k}) \,, 
\end{equation}
where $R^{(\ell)}_k=\int d \theta_{\hat k} R_{\mathbf{k}} Y_{\ell 0}(\theta_{\hat k})$.
Projecting the convolution term in Eq.~\eqref{eq:BS} on the spherical harmonics, we obtain
% \begin{widetext}
\begin{equation}
\begin{aligned}
 \int \frac{d\mathbf p}{(2\pi)^d} \Gamma_{|\mathbf{k}-\mathbf{p}|} F_{\mathbf p} &= \frac{\Omega_d}{2} \int \frac{d\mathbf p}{(2\pi)^d} G_p G_{-p} \times \\
& \sum_{\ell=0}^{\infty} \Gamma^{(\ell)}_{k,p}  Y_{\ell 0}(\theta_{\hat k})Y_{\ell 0}^{*}(\theta_{\hat p})   
 \sum_{\ell^\prime=0}^{\infty} R^{(\ell^\prime)}_k Y_{\ell^\prime 0}(\theta_{\hat k}) \\
&= \sum_{\ell=0}^{\infty} \frac{\Omega_d}{2}  \int \frac{p^{d-1}dp}{(2\pi)^d} \Gamma^{(\ell)}_{k,p} R^{(\ell)}_k Y_{\ell 0}(\theta_{\hat k}) \,,
\end{aligned}
\end{equation}
% \end{widetext}
where the orthonormal property of the spherical harmonics as $\int {\rm d}\theta_{\hat k} Y_{\ell m}(\theta_{\hat k})  Y^*_{\ell^\prime m^\prime}(\theta_{\hat k})  = \delta_{\ell \ell^\prime} \delta_{mm^\prime} $ is utilized.

Finally, we have the decoupled self-consistent linear response equation for each orbital channel in $d(\geq 3)$ dimensions as
\begin{equation}
R^{(\ell)}_k 
%= 1 -  \int \frac{p^2 {\rm d}p}{(2\pi)^2} \Gamma^{(\ell)}_{k,p} G_pG_{-p} R^{(\ell)}_p \,.
= 1 - \frac{\Omega_d}{2} \int \frac{p^{d-1} {\rm d}p}{(2\pi)^d} \Gamma^{(\ell)}_{k,p} G_pG_{-p} R^{(\ell)}_p \,,
\label{eq:dDR}
\end{equation}
where only the measure of momentum integral is different from two dimensions.
We can unify both Eq.~\eqref{eq:2DR} and Eq.~\eqref{eq:dDR} with the following simplified expression
\begin{eqnarray}
\label{eq:LR_expand}
R_K=1-\int dP \Gamma_{K,P} G^{(2)}_P R_P \,,
\end{eqnarray}
where we introduce the Capital-letter momentum-frequency variable $P=(\Delta p, \omega_m)$ with $\Delta p= p-\kf$ and $G^{(2)}_P \equiv G_PG_{-P}$. The integral is defined as $\int dP \equiv T\sum\limits_{m}\int d\Delta p$. We have also defined $\Gamma_{K,P}=\frac{p}{4\pi^2} \Gamma^{(\ell)}_{k,p}$ in two dimensions and 
%$\frac{p^2}{4\pi^2} \Gamma^{(\ell)}(k,\omega_n, p,\omega_m)$ in three dimensions 
$\frac{p^{d-1} \Omega_d}{2(2\pi)^d} \Gamma^{(\ell)}_{k,p}$ in higher dimensions
to absorb the measure of momentum integral. The $\ell$ label in $R^{(\ell)}_k$ is omitted.

\section{Superconducting phases of the two-dimensional electron gas with $G_0W_0$ self-energy renormalization}

\begin{figure}
\centering
\includegraphics[width=0.95\linewidth]{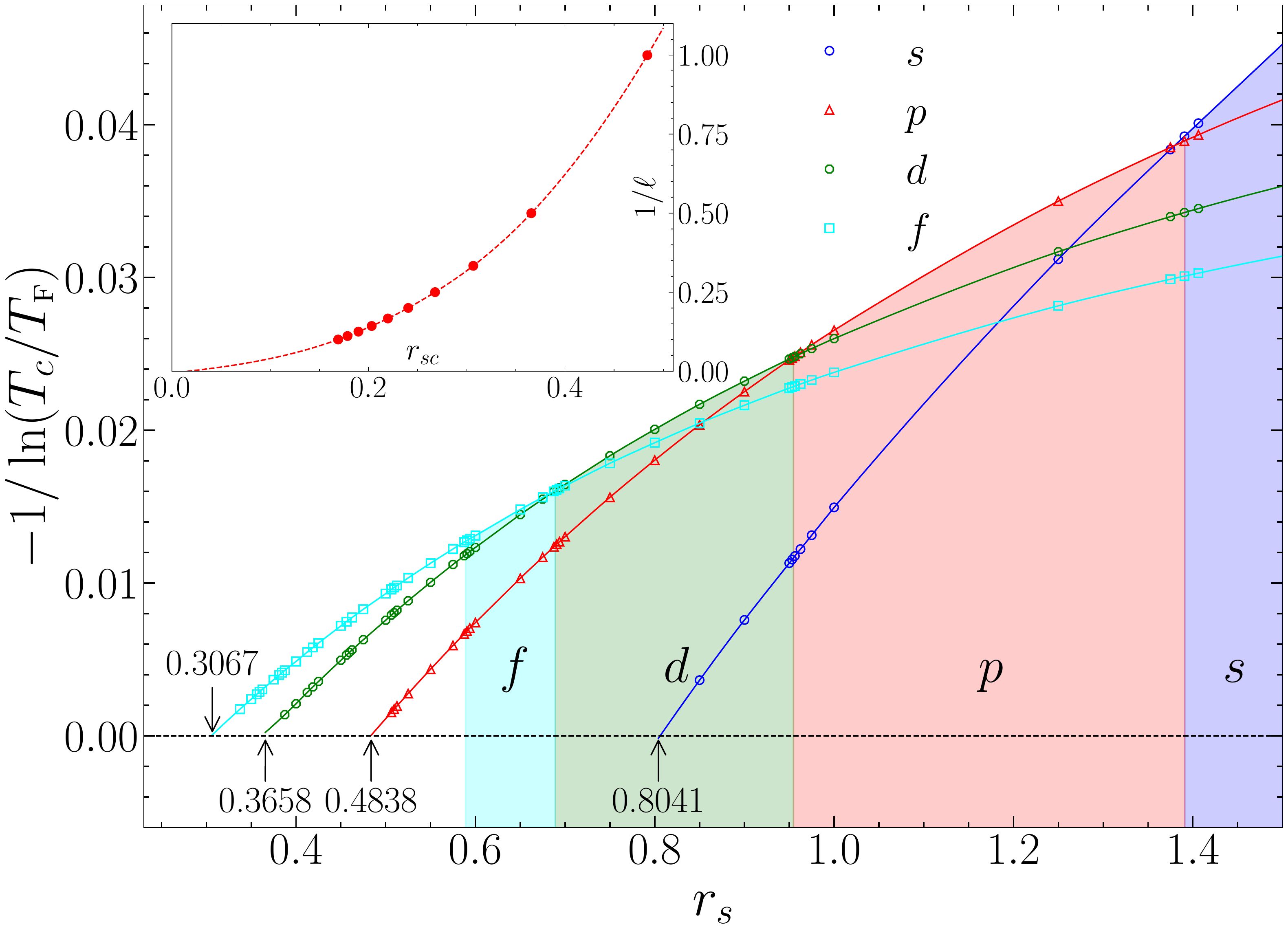}
\caption{Phase diagram of superconducting states in the 2D UEG within the 
RPA-$G^{\rm R}$ framework showing $s-$, $p-$, $d-$, and $f-$wave superconducting states. Each orbital channel features a QTP  with $T_c^{(\ell)} (r_s \to r_c^{(\ell)}) \to 0$. Critical values of $r_s$ for $\ell$ as large as $10$ are presented in the inset. 
}
\label{Fig:phase_rs} 
\end{figure}

We employed the precursory Cooper flow approach to investigate dynamic-screening-driven superconductivity in the two-dimensional uniform electron gas (2D UEG) model, parametrized by the Wigner-Seitz radius $r_s = \sqrt{2}/(k_{\rm F}a_{\rm B})$, where $k_{\rm F}$ is the Fermi momentum and $a_{\rm B}$ is the Bohr radius. As shown in Fig.~1 of the main text, we obtained a rich phase diagram for several channels.

To verify the robustness of the pairing mechanism in the 2D UEG, here, we computed the same phase diagram using the RPA interaction for $\Gamma$ and the $G_0W_0$ renormalized propagator $G^{\textsc r}$ (referred to as RPA-$G^{\rm R}$), as shown in Fig.~\ref{Fig:phase_rs}.
%and compared it with the results from RPA-$G_0$ presented in the main text. Despite the shifted phase boundaries, 
Comparison with the RPA-$G_0$ results presented in the main text reveals small shifts of the phase boundaries, but all key features of the RPA-$G_0$ phase diagrams are preserved for RPA-$G^{\rm R}$. Namely: (i) The critical temperature $T_c$ in all orbital channels is suppressed as density increases ($r_s$ decreases) and eventually becomes zero at the quantum transition point (QTP) $r_c^{(\ell)}$, below which the 2D UEG no longer superconducts. (ii) On the logarithmic scale, $-1/\ln [T_c(r_s)/T_{\rm F}]$ still demonstrates a linear dependence on $r_s$ near the QTP, indicating that the BCS transition temperature obeys $T_c^{(\ell)}=T_{\textsc f}e^{-1/\lambda_\ell}$ with the coupling parameter $\lambda_\ell = (r_s-r_c^{(\ell)})/c_{\ell}$, where $c_{\ell}$ is a constant. (iii) The data for $1/\ell$ versus $r_c$ in the inset of Fig.~\ref{Fig:phase_rs} suggests that superconductivity survives in the high-density limit for large enough $\ell$. The qualitative agreement between RPA-$G_0$ and RPA-$G^{\rm R}$ phase diagrams reflects the robustness of dynamic-screening-driven superconductivity in the 2D UEG.

\section{Superconducting phases of the two-dimensional electron gas with Yukawa-type interaction}

To investigate the impact of interaction range on the dynamic-screening-driven superconductivity in two dimensions, we consider the 2D electron gas with the Yukawa-type interaction $V_{\rm Y}(q)=2\pi e^2/\sqrt{q^2+q_0^2}$, also known as a statically screened Coulomb interaction, where $q_0$ is the screening momentum. Specifically, we examine the behavior of the superconducting state in the 2D electron gas at $r_s=0.8$ when all channels exhibit superconductivity in the Coulomb system. We performed systematic and extensive calculations of the linear response function $R$ for various channels $\ell$ and values of $q_0$ using RPA-$G_0$ and RPA-$G^{\rm R}$, respectively.

\begin{figure*}
	\centering
    \includegraphics[width=0.45\linewidth]{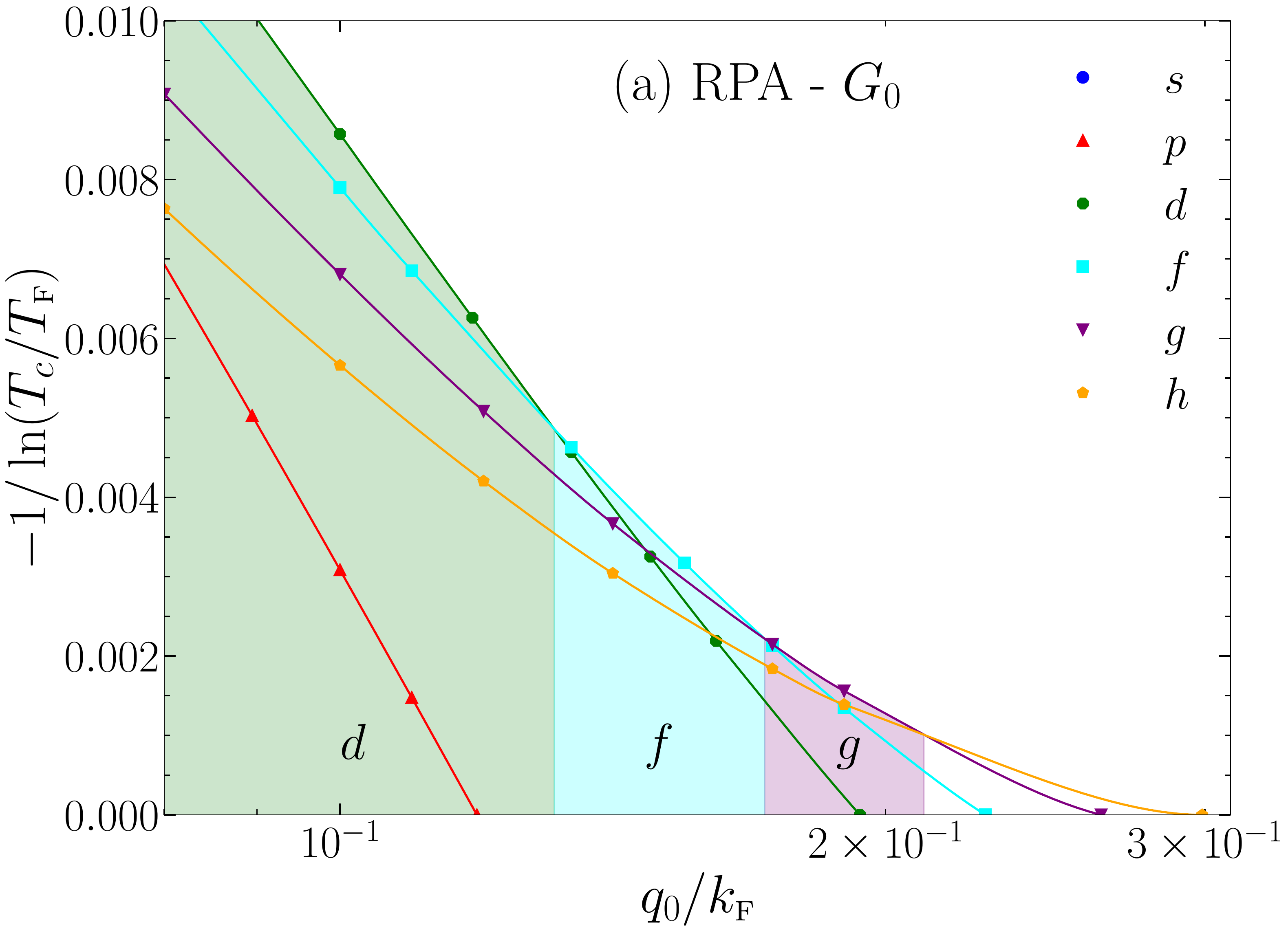}	\includegraphics[width=0.45\linewidth]{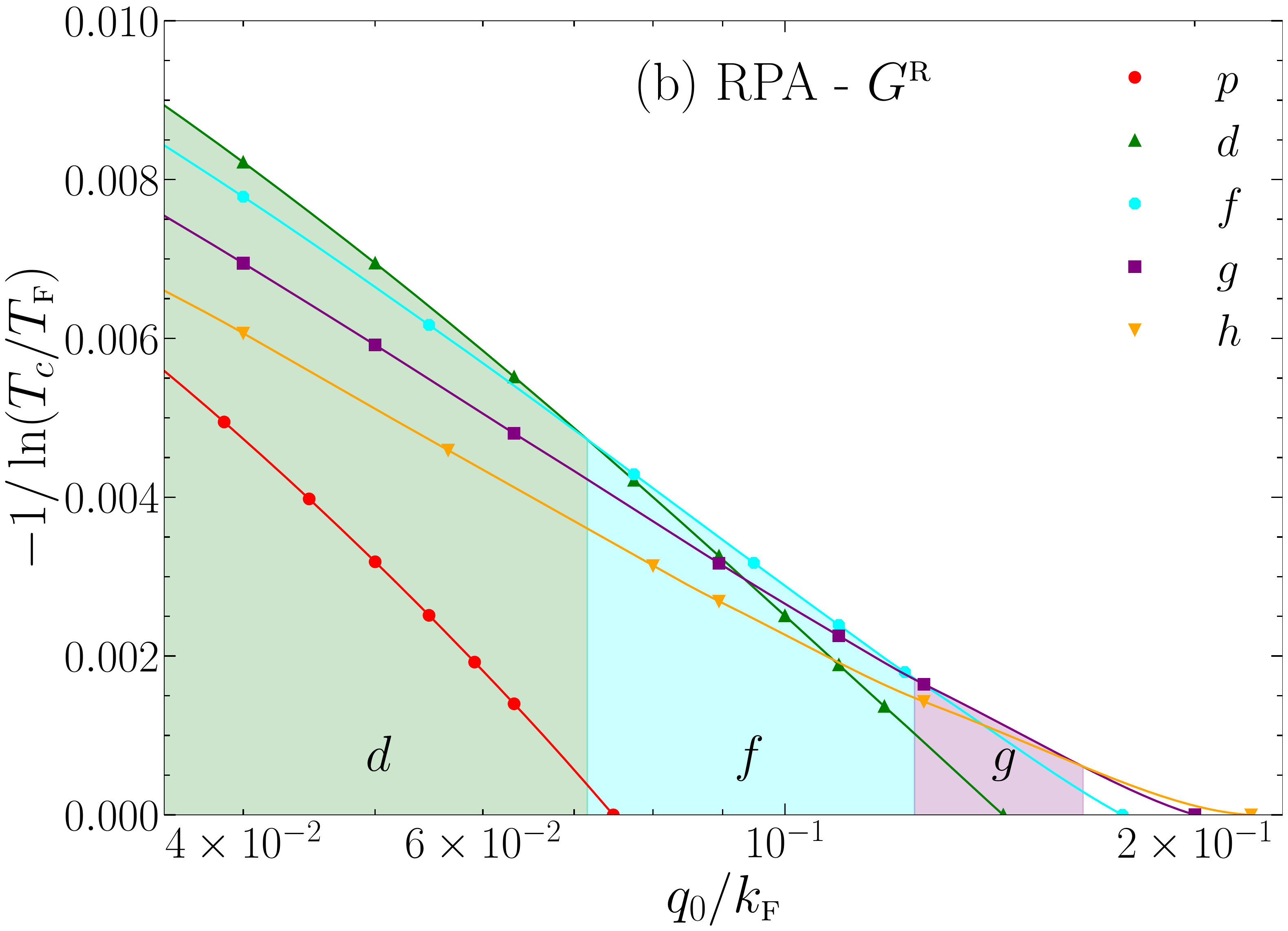}
	\caption{
    Ultralow-temperature phase diagram of a 2D electron gas with $r_s=0.8$ and a Yukawa-type interaction $V_{\rm Y}(q)=2\pi e^2/\sqrt{q^2+q_0^2}$, calculated using the (a) RPA-$G_0$ and (b) RPA-$G^{\rm R}$ approximations.  
    The superconducting temperature ($T_c$) as a function of the screening momentum ($q_0$) is represented by lines going through data points for different channels, with the colored shadow indicating the dominant channel. For any given channel $\ell$, $T_c^{(\ell)}$ decreases with increasing $q_0$ until it disappears at the critical value $q_{0c}^{(\ell)}$.
    }
	\label{Fig:phase_q0} 
\end{figure*}

For each $\ell$ and $q_0$, we determined the transition temperature $T_c$ by extrapolating the precursory Cooper flow to $1/R_0(T_c)=0$. Subsequently, we used extrapolation to determine the critical values of $q_0$ ($q_{0c}$), where $-1/\ln[T_c(q_0)/T_{\rm F}]$ becomes zero and superconductivity vanishes. Finally, we obtained the phase diagrams in the 2D electron gas with the Yukawa-type interaction under the RPA-$G_0$ [Fig.~\ref{Fig:phase_q0}(a)] and RPA-$G^{\rm R}$ [Fig.~\ref{Fig:phase_q0}(b)] approximations, respectively. The qualitative agreement between the two phase diagrams indicates that the approximations are controlled.
	  
As the range of interaction decreases (i.e., as $q_0$ increases), the superconducting critical temperature $T_c$ decreases for any given channel $\ell$ and eventually drops to zero at the QTP with the critical screening momentum $q_{0c}^{(\ell)}$. Since the rate of $T_c$ drop slows as $\ell$ increases, the dominant superconducting channel, determined from the crossing point of $T_c^{(\ell)}(q_0)$ and $T_c^{(\ell+1)}(q_0)$, increases as $q_0$ increases. 
For $q_0 > 0.2 k_{\rm F}$, all orbital channels shown in Fig.~\ref{Fig:phase_q0} are no longer superconducting!
These observations suggest that the long-range character of the Coulomb interaction is essential in dynamic-screening-driven superconductivity and that decreasing the interaction range suppresses superconductivity in the 2D electron gas. Furthermore, for $q_0 \to q_{0c}^-$, a linear relation exists between the inverse logarithmic scale $-1/\ln [T_c(q_0)/T_{\rm F}]$ and $q_0$, indicating that the superconductivity near the QTP can be described by the emergent BCS theory with a coupling parameter $\lambda^\prime_{\ell} \propto (q_{0c}-q_0)/\kf \ll 1$.

\section{Data table of quantum transition points}
Results for quantum transition points $r_c^{(\ell)}$ and left boundaries of the dominant superconducting channel $r_{st}(\ell \to \ell+1)$ for various orbital 
channels $\ell$ are given in Tab.~\ref{tab:rsct}. The data for 
dimensionless critical screening momentum $q_{0c}^{(\ell)}$ at $r_s=0.8$ are given in Tab.~\ref{tab:msc}.
\begin{table}[]
    \centering
    \begin{tabular}{|c|c|c|c|c|}
    \hline
      $\ell$  & $r_c^{{\rm RPA-}G_0}$ & $r_c^{{\rm RPA-}G^{\textsc r}}$ & $r_{st}^{{\rm RPA-}G_0}(\ell \to \ell+1)$ & $r_{st}^{{\rm RPA-}G^{\textsc r}}$  \\
      \hline
     0 & 0.6339(12) & 0.8041(25) & 1.156(3)  & 1.391(16)\\
     1 & 0.4196(10) & 0.4838(20) & 0.8484(16)  & 0.9547(16)\\
     2 & 0.3287(10) & 0.3658(17) & 0.6234(16)  & 0.6891(16)\\
     3 & 0.2806(10) & 0.3067(11) & 0.5359(16)  & 0.5891(16)\\
     4 & 0.2479(6) & 0.2678(11) & 0.4672(16)  & 0.5078(16)\\
     5 & 0.2244(7) & 0.2405(18) & 0.4234(16)  & 0.4578(16)\\
     6 & 0.2062(5) & 0.2198(20) & 0.3859(16)  & 0.416(3)\\
     7 & 0.1918(8) & 0.2033(10) & 0.3578(16)  & 0.3828(16)\\
     8 & 0.1798(5) & 0.1898(10) & 0.3359(16)  & 0.3578(16)\\
     9 & 0.1698(4) & 0.1787(10) & 0.3141(16)  & 0.331(6)\\
    10 & 0.1611(5) & 0.1692(14) &-         & -\\
    \hline
    \end{tabular}
    \caption{The critical $r_s$ parameters for the 2D electron gas under the RPA-$G_0$ and RPA-$G^{\textsc r}$ approximations determine the disappearance of superconductivity in channel $\ell$ at $r_c$ and the transition from the dominant channel $\ell$ to $\ell+1$ at $r_{st}$.}
    \label{tab:rsct}
\end{table}

\begin{table}[]
    \centering
    \begin{tabular}{|c|c|c|c|}
    \hline
      $r_s$ & $\ell$ & $q_{0c}^{{\rm RPA-}G_0}$  & $q_{0c}^{{\rm RPA-}G^{\textsc r}}$  \\
      \hline
     & 2 & 0.00674(15)  & 0.00152(7) \\
     & 3 & 0.0193(2)  & 0.0110(3) \\
     & 4 & 0.0313(3)  & 0.0216(5) \\
     & 5 & 0.0409(4)  & 0.0306(7) \\
    0.4 & 6 & 0.0490(5)  & 0.0382(8) \\
     & 7 & 0.0557(6)  & 0.0445(9) \\
     & 8 & 0.0615(6)  & 0.0500(10) \\
     & 9 & 0.0666(6)  & 0.0546(11) \\
     & 10 & 0.0713(7)  & 0.0588(13) \\
     \hline
    & 0  & 0.0205(5) & -  \\
    & 1 & 0.1190(11) & 0.0748(7) \\
    & 2 & 0.1936(13) & 0.1446(18) \\
0.8 & 3 & 0.2270(11) & 0.177(6) \\
    & 4 & 0.263(4) & 0.205(5) \\
    & 5 & 0.299(7) & 0.228(9) \\
      \hline
    \end{tabular}
    \caption{Critical screening momentum $q_{0c}$ (in units of the Fermi momentum $\kf$) of the 2D electron gas with a Yukawa-type interaction under the RPA-$G_0$ and RPA-$G^{\textsc r}$ approximations at $r_s=0.4$ and 0.8. Note that $s$-wave superconductivity is absent in the 2D electron gas with RPA-$G^{\textsc r}$ at $r_s=0.8$ and $q_{0} = 0$. }
    \label{tab:msc}
\end{table}

\section{Data table of superconducting temperatures of the two-dimensional electron gas}

All the superconducting transition temperatures $T_c^{(\ell)}$ data are given in Tab.~\ref{tab:Tc_G0} with RPA-$G_0$ and Tab.~\ref{tab:Tc_GR} with RPA-$G^{\textsc r}$.

\clearpage
\begin{center}
\tablefirsthead{%
\hline
 $r_s$ & $T_c^{(0)}$ & $T_c^{(1)}$  &$T_c^{(2)}$ &$T_c^{(3)}$ \\
 \hline}
%  \tablehead{%
%  \hline
% 	\multicolumn{5}{l}{\small\sl continued from previous page}\\
% \hline
%  $r_s$ & $T_c^{(0)}$ & $T_c^{(1)}$  &$T_c^{(2)}$ &$T_c^{(3)}$ \\
%  \hline}
%  \tabletail{%
% 	\hline
% 	\multicolumn{5}{r}{\small\sl continued on next page}\\
% 	\hline}
 \tablelasttail{\hline}
\topcaption{\label{tab:Tc_G0} Superconducting temperatures $T_c^{(\ell)}$ (in units of the Fermi temperature $T_{\rm F}$) for different $r_s$ from RPA-$G_0$. The `-' symbol means that the $T_c$ value is lower than the limit of double-precision floating-point number resolution.}
\begin{supertabular}{c@{\hspace{2mm}} c@{\hspace{3mm}} c@{\hspace{3mm}} c@{\hspace{3mm}} c@{\hspace{3mm}}}

0.300000 &0 &0 &0 &1.406e-237 \\  
0.312500 &0 &0 &0 &4.072e-147 \\
0.315625 &0 &0 &0 &2.579e-134 \\
0.318750 &0 &0 &0 &1.469e-123 \\
0.325000 &0 &0 &0 &1.695e-106 \\
0.331250 &0 &0 &-& 1.4249e-93 \\
0.334375 &0 &0 &-& 3.2153e-88 \\
0.337500 &0 &0 &-& 1.9099e-83 \\
0.350000 &0 &0 &1.280e-189& 1.3299e-68 \\
0.356250 &0 &0 &2.575e-147& 5.2761e-63 \\
0.359375 &0 &0 &1.572e-132& 1.5520e-60 \\
0.362500 &0 &0 &1.927e-120& 2.9667e-58 \\
0.375000 &0 &0 &2.9758e-88 &1.2427e-50 \\  
0.381250 &0 &0 &7.3936e-78 &1.5765e-47 \\
0.384375 &0 &0 &1.5728e-73 &4.0684e-46 \\
0.387500 &0 &0 &1.1665e-69 &8.6837e-45 \\
0.400000 &0 &0 &1.4701e-57 &3.6303e-40 \\
0.412500 &0 &0 &4.6208e-49 &2.0135e-36 \\
0.418750 &0 &0 &1.0688e-45 &8.3380e-35 \\
0.421875 &0 &-&3.4799e-44 &4.7395e-34 \\
0.425000 &0 &-&9.0343e-43 &2.4979e-33 \\
0.450000 &0 &2.007e-112 &4.3625e-34 &1.6146e-28 \\
0.462500 &0 &1.2949e-79 &5.7451e-31 &1.2998e-26 \\
0.465625 &0 &3.2625e-74 &2.8157e-30 &3.5458e-26 \\
0.468750 &0 &1.6743e-69 &1.2848e-29 &9.3524e-26 \\
0.475000 &0 &1.0989e-61 &2.1986e-28 &5.9198e-25 \\
0.500000 &0 &1.4072e-42 &2.0830e-24 &2.8734e-22 \\
0.525000 &0 &1.1742e-32 &2.2786e-21 &4.9251e-20 \\
0.531250 &0 &7.2448e-31 &9.7356e-21 &1.4711e-19 \\
0.534375 &0 &4.8022e-30 &1.9460e-20 &2.4908e-19 \\
0.537500 &0 &2.8769e-29 &3.8086e-20 &4.1629e-19 \\
0.550000 &0 &1.5595e-26 &4.6101e-19 &2.8777e-18 \\
0.600000 &0 &2.0370e-19 &9.6175e-16 &1.4137e-15 \\
0.618750 &0 &1.1135e-17 &8.4689e-15 &8.9310e-15 \\
0.621875 &0 &2.0163e-17 &1.1839e-14 &1.1902e-14 \\
0.625000 &0 &3.5848e-17 &1.6431e-14 &1.5776e-14 \\
0.650000 &6.976e-162 &2.0240e-15 &1.7896e-13 &1.2623e-13 \\
0.700000 &2.0992e-37 &7.2375e-13 &7.9388e-12 &3.7862e-12 \\
0.750000 &1.3675e-21 &4.2496e-11 &1.4054e-10 &5.4233e-11 \\
0.800000 &2.1308e-15 &8.3996e-10 &1.3343e-09 &4.6029e-10 \\
0.825000 &1.5734e-13 &2.8191e-09 &3.4548e-09 &1.1534e-09 \\
0.837500 &9.0592e-13 &4.8863e-09 &5.3631e-09 &1.7689e-09 \\
0.843750 &2.0085e-12 &6.3537e-09 &6.6275e-09 &2.1747e-09 \\
0.846875 &2.9382e-12 &7.2240e-09 &7.3530e-09 &2.4070e-09 \\
0.850000 &4.2506e-12 &8.1979e-09 &8.1475e-09 &2.6612e-09 \\
0.900000 &4.7595e-10 &4.9371e-08 &3.5988e-08 &1.1516e-08 \\
0.950000 &1.1847e-08 &2.1071e-07 &1.2450e-07 &3.9840e-08 \\
1.000000 &1.1952e-07 &6.8526e-07 &3.5015e-07 &1.1361e-07 \\
1.125000 &5.1861e-06 &6.5649e-06 &2.7484e-06 &9.4091e-07 \\
1.156250 &1.0027e-05 &1.0202e-05 &4.1536e-06 &1.4455e-06 \\
1.187500 &1.8013e-05 &1.5295e-05 &6.0903e-06 &2.1568e-06 \\
1.218750 &3.0440e-05 &2.2195e-05 &8.6879e-06 &3.1285e-06 \\
1.250000 &4.8980e-05 &3.1356e-05 &1.2097e-05 &4.4263e-06 \\
% \hline
    \end{supertabular}
% \end{table}    
\end{center}

\clearpage
\begin{center}
\tablefirsthead{%
\hline
 $r_s$ & $T_c^{(0)}$ & $T_c^{(1)}$  &$T_c^{(2)}$ &$T_c^{(3)}$ \\
 \hline}
%  \tablehead{%
%  \hline
% 	\multicolumn{5}{l}{\small\sl continued from previous page}\\
% \hline
%  $r_s$ & $T_c^{(0)}$ & $T_c^{(1)}$  &$T_c^{(2)}$ &$T_c^{(3)}$ \\
%  \hline}
%  \tabletail{%
% 	\hline
% 	\multicolumn{5}{r}{\small\sl continued on next page}\\
% 	\hline}
 \tablelasttail{\hline}
 \topcaption{\label{tab:Tc_GR} Superconducting temperatures $T_c^{(\ell)}$ (in units of the Fermi temperature $T_{\rm F}$) for different $r_s$ from RPA-$G^{\textsc r}$. The symbol `-' indicates that the value of $T_c$ is too small to be represented by a double-precision floating-point number.}
\begin{supertabular}{c@{\hspace{2mm}} c@{\hspace{3mm}} c@{\hspace{3mm}} c@{\hspace{3mm}} c@{\hspace{3mm}}}

0.337500 &0 &0 &0 &1.848e-248 \\
0.350000 &0 &0 &0 &2.831e-181 \\
0.356250 &0 &0 &0 &5.646e-160 \\
0.359375 &0 &0 &0 &3.665e-151 \\
0.362500 &0 &0 &0 &2.615e-143 \\
0.375000 &0 &0 &-&6.452e-119 \\
0.381250 &0 &0 &-&1.003e-109 \\
0.384375 &0 &0 &-&1.130e-105 \\
0.387500 &0 &0 &1.366e-313&6.269e-102 \\
0.400000 &0 &0 &2.412e-207& 6.6615e-90 \\
0.412500 &0 &0 &3.464e-153& 7.7736e-80 \\
0.418750 &0 &0 &3.555e-136& 7.7817e-76 \\
0.425000 &0 &0 &1.111e-122& 2.9707e-72 \\
0.450000 &0 &0 &1.8986e-88 &4.9793e-61 \\
0.456250 &0 &0 &8.2578e-83 &8.3200e-59 \\
0.459375 &0 &0 &2.8627e-80 &9.1969e-58 \\
0.462500 &0 &0 &6.8269e-78 &9.2354e-57 \\
0.475000 &0 &0 &9.8906e-70 &3.9974e-53 \\
0.500000 &0 &-&4.3555e-58 &2.2543e-47 \\
0.506250 &0 &8.491e-282&1.2004e-55 &5.0364e-46 \\
0.509375 &0 &9.489e-249&1.4192e-54 &1.9576e-45 \\
0.512500 &0 &9.849e-223&1.5107e-53 &7.3012e-45 \\
0.525000 &0 &1.667e-157&7.6809e-50 &9.6863e-43 \\
0.550000 &0 &3.180e-100&6.1712e-44 &3.7644e-39 \\
0.575000 &0 &3.8725e-74 &1.9191e-39 &3.1175e-36 \\
0.587500 &0 &8.6714e-66 &1.4080e-37 &5.7148e-35 \\
0.590625 &0 &5.2670e-64 &3.8236e-37 &1.1358e-34 \\
0.593750 &0 &2.5354e-62 &1.0101e-36 &2.2238e-34 \\
0.600000 &0 &2.6681e-59 &5.8416e-36 &7.2998e-34 \\
0.650000 &0 &7.7443e-43 &1.0118e-30 &4.8374e-30 \\
0.675000 &0 &7.4252e-38 &9.2698e-29 &1.5266e-28 \\
0.687500 &0 &7.9884e-36 &6.8043e-28 &7.2234e-28 \\
0.690625 &0 &2.3548e-35 &1.0932e-27 &1.0485e-27 \\
0.693750 &0 &6.7210e-35 &1.7405e-27 &1.5126e-27 \\
0.700000 &0 &4.7454e-34 &4.0702e-27 &2.9090e-27 \\
0.750000 &0 &1.5585e-28 &2.0370e-24 &4.5382e-25 \\
0.800000 &0 &8.5340e-25 &2.2287e-22 &2.3030e-23 \\
0.850000 &9.797e-120&4.6114e-22 &9.6861e-21 &6.0225e-22 \\
0.900000 &5.7897e-58 &5.2413e-20 &1.9609e-19 &8.5010e-21 \\
0.950000 &3.8577e-39 &2.2112e-18 &2.4495e-18 &8.2838e-20 \\
0.953125 &2.2514e-38 &2.7177e-18 &2.8234e-18 &9.4232e-20 \\
0.956250 &1.2216e-37 &3.3309e-18 &3.2494e-18 &1.0706e-19 \\
0.962500 &2.9409e-36 &4.9637e-18 &4.2845e-18 &1.3767e-19 \\
0.975000 &8.4522e-34 &1.0689e-17 &7.3212e-18 &2.2440e-19 \\
1.000000 &9.0407e-30 &4.3732e-17 &1.9723e-17 &5.5433e-19 \\
1.250000 &1.1905e-14 &3.8819e-13 &1.9591e-14 &3.7003e-16 \\
1.375000 &4.8188e-12 &5.2926e-12 &1.6566e-13 &2.9358e-15 \\
1.390625 &8.5285e-12 &6.9704e-12 &2.0832e-13 &3.6727e-15 \\
1.406250 &1.4732e-11 &9.0872e-12 &2.6001e-13 &4.5643e-15 \\
\hline
    \end{supertabular}
% \end{table}    
\end{center}

\bibliographystyle{apsrev}
\bibliography{references}